\documentclass[prx,twocolumn,amsmath,amssymb,aps,longbibliography,superscriptaddress,citeautoscript,notitlepage,bibnotes,nofootinbib,floatfix]{revtex4-2}
\usepackage{amsmath,amssymb}
\usepackage{bm}
\usepackage{graphicx}
\usepackage{braket}
\usepackage{booktabs}
\usepackage[normalem]{ulem}
\usepackage[dvipsnames]{xcolor}
\usepackage{tikz}
\usepackage[colorlinks,bookmarks=true,citecolor=blue,linkcolor=red,urlcolor=blue]{hyperref}
\usepackage{tcolorbox}
\newtcolorbox{resultbox}[1]{colback=white,colframe=black!55,boxrule=0.5pt,arc=0pt,left=4pt,right=4pt,top=3pt,
  bottom=3pt,toptitle=1pt,bottomtitle=1pt,fonttitle=\bfseries,coltitle=black,colbacktitle=black!8,title={#1},
  before skip=8pt,after skip=8pt}

\newcommand{\acoh}{\alpha_{\rm coh}}
\newcommand{\fs}{f_*}
\newcommand{\CCZ}{\mathrm{CCZ}}
\newcommand{\Mlin}{M_3^{\rm lin}}
\newcommand{\QI}{\mathbb{Q}(i)}

\renewcommand{\eqref}[1]{\ref{#1}}

\begin{document}
\title{Magic of Dicke states: density, cost and yield}

\date{\today}

\author{Yves H. Kwan} \affiliation{Department of Physics, University of Texas at Dallas, Richardson, Texas 75080, USA}

\begin{abstract}
Dicke states are an elementary family of many-body quantum states, consisting of equal superpositions of all configurations of $L$ qubits with $n$ excitations. We show that their magic, as measured by the stabilizer R\'enyi entropy (SRE), is extensive at every filling $f=n/L$ except half-filling, and we obtain the magic density analytically, together with the subleading terms. 
In the context of fault-tolerant quantum computation, we also investigate the yield of Dicke states, and find that they can produce on average an extensive number of CCZ states (or $T$ states with a catalyst) at any $0<f<1$. This yield is not self-averaging, but fluctuates from run to run by $O(L)$, as if it were set by a hidden phase shared by all qubits. At half-filling, the SRE grows only logarithmically, while the preparation cost, average yield and relative entropy of magic remain extensive, so that a small SRE does not necessarily signal a weak resource. These results can be understood by viewing the Dicke state as a fixed-magnetization projection of a product of tilted spins.

\end{abstract}

\maketitle

\section{Introduction}\label{sec:intro}

Dicke states~\cite{Dicke1954coherence} are the equal-weight superpositions of all configurations of $L$ spin-$\frac{1}{2}$ degrees of freedom with $n$ flipped spins. They have been experimentally prepared in platforms based on photons and trapped ions~\cite{Kiesel2007experimental,Prevedel2009experimental,Wieczorek2009experimental,Hume2009preparation,Roos2004control,Haffner2005scalable}, and are standard targets of state-preparation algorithms~\cite{Bartschi2019deterministic,Yu2024efficient,Joshi2026constant} and resources for quantum metrology~\cite{Pezze2018quantum,Lucke2011twin,Zou2018beating}. 
Dicke states also arise as the ground states of the Heisenberg ferromagnet at fixed magnetization, and as exact towers of eigenstates in models of quantum many-body scars and $\eta$-pairing~\cite{Choi2018emergent,Serbyn2021quantum,Moudgalya2022quantum,Chandran2023quantum,Yang1989eta}.

Quantum circuits consisting of Clifford gates, stabilizer states, and Pauli measurements can be simulated efficiently on a classical computer~\cite{Gottesman1998heisenberg,Aaronson2004improved}. In fault-tolerant quantum computation, non-Clifford ingredients are therefore considered expensive, and are supplied by special resource states, most notably $T$ and CCZ states. The latter are typically prepared by magic-state distillation~\cite{Bravyi2005universal} and consumed by gate teleportation~\cite{Gottesman1999demonstrating}. The non-Clifford character of a state is called non-stabilizerness, also known as magic~\cite{Veitch2014resource,Howard2017application}. For many-qubit states, magic is often quantified by the (second) stabilizer R\'enyi entropy (SRE) $M_2$~\cite{leone2022SRE,Oliviero2022magicstate,Haug2022quantifying,Lami2023nonstabilizerness,Tarabunga2023manybody,Haug2023scalable}.

The SRE of Dicke states has so far been studied in two regimes. For a fixed number $n$ of excitations, $M_2$ grows only logarithmically with $L$, as known in closed form for the $W$ state ($n=1$)~\cite{Odavic2022complexity,Catalano2024magic} and for $n=2,3$~\cite{Liu2025stabilizer}. At half-filling $f=1/2$, a large-$L$ analysis gives $M_2\simeq2\log_2(\pi L/8)-1$~\cite{HernandezYanes2025non}. Exact expressions for every $L$ and $n$ are available as finite binomial sums~\cite{Liu2026quantifying,Liu2026quantifyingcodes}; while these can be evaluated numerically, they do not directly probe the scaling of magic in the thermodynamic limit at fixed filling. Extensive magic has been found in other permutation-symmetric states~\cite{passarelli2024permutation}, and in typical states of a fixed $U(1)$ charge sector~\cite{Iannotti2026non}. However, the magic density and resource-theoretic properties of Dicke states at generic fillings have yet to be addressed.

In this work, we demonstrate analytically that the SRE of a Dicke state is generally extensive, $M_2\simeq\alpha(f)L$, 
with the magic density $\alpha(f)$ obtained in closed form. The density vanishes at half-filling, while its largest value, $\log_2\frac{4}{3}$ bits per qubit, equals the magic of one $T$ state. We further derive the subleading logarithmic correction $2\log_2L$ and the constant term $c(f)$. Turning to resource theory, we show that preparing a Dicke state costs at least about one $T$ state per qubit. 
Moreover, heralded procedures~\cite{Rizzo2026universal} extract on average an extensive number of CCZ states (or $T$ states if one catalyst $T$ is introduced) for $0<f<1$. Interestingly, the yield does not self-average, and is distributed according to a hidden phase shared by all qubits. The magic density and the CCZ yield can both be understood through the common lens of a Dicke state as a product of tilted spins, projected onto a fixed number of excitations by restoring the $U(1)$ symmetry (Fig.~\ref{fig:proj}). 
This projection picture also explains the intriguing finding that the half-filling Dicke state has vanishing magic density, yet provides on average an extensive number of resource states. We introduce the projection picture and compute the magic density in Sec.~\ref{sec:density}, discuss the cost and yields of Dicke states in Sec.~\ref{sec:usefulness}, and close with an outlook in Sec.~\ref{sec:outlook}.

\section{Magic density}\label{sec:density}

\subsection{Dicke states and the stabilizer R\'enyi entropy}\label{subsec:defs}

We consider a system of $L$ qubits. In the spin-$\frac{1}{2}$ language, the up (down) state of a qubit is $\ket{0}$ ($\ket{1}$), and each $\ket{1}$ corresponds to an excitation. A Dicke state is the equal-amplitude superposition of all configurations with $n$ excitations,
\begin{equation}\label{eq:dicke}
\ket{D^L_n}=\binom{L}{n}^{-1/2}\sum_{x\in\{0,1\}^L,\ |x|=n}\ket{x},
\end{equation}
where $|x|$ counts the number of 1's in the bit string $x$. $\ket{D^L_n}$ is the $S^z=\frac{1}{2}\sum_jZ_j=L/2-n$ member of the multiplet with total spin $L/2$, where $Z_j$ is the Pauli $Z$ operator on qubit $j$. Flipping every spin with $X^{\otimes L}$ maps $\ket{D^L_n}$ to $\ket{D^L_{L-n}}$, so it suffices to consider fillings $f=n/L\le\frac{1}{2}$.

A Pauli string $P$ on $L$ qubits is a tensor product of single-qubit operators $I,X,Y,Z$. A stabilizer state is a common eigenstate, with eigenvalues $\pm1$, of $L$ independent commuting Pauli strings. Equivalently, a stabilizer state can be obtained from $\ket{0}^{\otimes L}$ by a Clifford unitary, which maps Pauli strings to each other under conjugation. 
The Clifford unitaries are generated by the Hadamard gate $H$, the phase gate $S=\mathrm{diag}(1,i)$, and the two-qubit CNOT gate.

For any pure state, $\langle P\rangle^2/2^L$ is a normalized probability distribution over the $4^L$ Pauli strings, known as the Pauli spectrum of the state. A stabilizer state concentrates its Pauli spectrum on $2^L$ strings with $\langle P\rangle=\pm1$, whereas a magical state can distribute it over many strings with small expectation values. The (second) SRE~\cite{leone2022SRE} quantifies this spread,
\begin{equation}\label{eq:M2}
M_2(\psi)=-\log_2\Big(2^{-L}\sum_P\langle P\rangle^4\Big),
\end{equation}
which is non-negative, additive, invariant under Clifford unitaries, and vanishes for stabilizer states. 

For a single qubit with Bloch vector $\mathbf{r}$,
\begin{equation}\label{eq:M2qubit}
M_2=-\log_2\frac{1+r_x^4+r_y^4+r_z^4}{2},
\end{equation}
which vanishes on the six single-qubit stabilizer states. The $T$ state $\ket{T}=(\ket{0}+e^{i\pi/4}\ket{1})/\sqrt2$ has $M_2=\log_2\frac{4}{3}$, and the most magical single-qubit state, whose Bloch vector points along $(1,1,1)/\sqrt3$, has $M_2=\log_2\frac{3}{2}$. 
Additivity makes it natural to define the magic density as the magic per qubit in the thermodynamic limit,
\begin{equation}\label{eq:alphadef}
\alpha(f)=\lim_{L\to\infty}\frac{M_2(\ket{D^L_n})}{L},
\end{equation}
where the limit is taken at fixed filling $f=n/L$.

\subsection{A Dicke state as a projected product state}\label{subsec:mechanism}

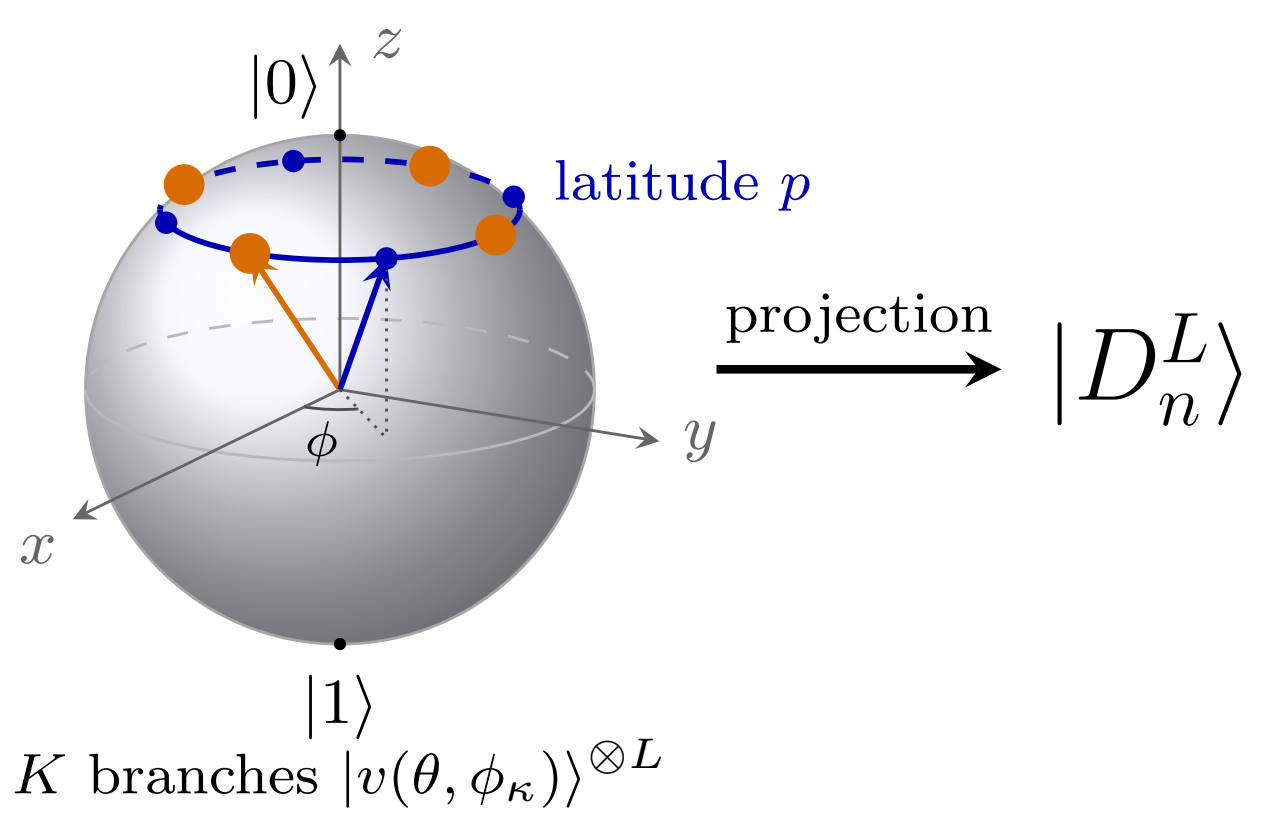
\begin{figure}[t]
\centering
\resizebox{\columnwidth}{!}{\begin{tikzpicture}[x=1cm,y=1cm,>=stealth,font=\small]
  \def\R{1.25}
  \def\th{45}       
  \def\xang{-120}      
  \pgfmathsetmacro{\yc}{\R*cos(\th)}
  \pgfmathsetmacro{\rx}{\R*sin(\th)}
  \pgfmathsetmacro{\ry}{0.28*\R*sin(\th)}
  \shade[ball color=blue!4, opacity=0.9] (0,0) circle (\R);
  \draw[gray!70] (0,0) circle (\R);
  \draw[gray!55, dashed] (\R,0) arc (0:180:\R cm and 0.28*\R cm);
  \draw[gray!55] (-\R,0) arc (180:360:\R cm and 0.28*\R cm);
  \draw[->, gray!80!black] (0,0) -- (0,\R+0.45) node[right=1pt] {$z$};
  \draw[->, gray!80!black] (0,0) -- ({2.1*\R*cos(\xang)},{2.1*0.28*\R*sin(\xang)}) node[below left=-1pt] {$x$};
  \draw[->, gray!80!black] (0,0) -- ({1.45*\R*cos(\xang+90)},{1.45*0.28*\R*sin(\xang+90)}) node[right=0pt] {$y$};
  \fill (0,\R) circle (0.03) node[above left=-1pt] {$\ket{0}$};
  \fill (0,-\R) circle (0.03) node[below=1pt] {$\ket{1}$};
  \draw[blue!70!black, dashed, thick] (\rx,\yc) arc (0:180:\rx cm and \ry cm);
  \draw[blue!70!black, thick] (-\rx,\yc) arc (180:360:\rx cm and \ry cm);
  \draw[->, orange!85!black, thick] (0,0) -- ({\rx*cos(\xang)},{\yc+\ry*sin(\xang)});
  \draw[->, blue!70!black, thick] (0,0) -- ({\rx*cos(\xang+45)},{\yc+\ry*sin(\xang+45)});
  \draw[gray!60!black, densely dotted] ({\rx*cos(\xang+45)},{\yc+\ry*sin(\xang+45)}) -- ({\rx*cos(\xang+45)},{0.28*\rx*sin(\xang+45)});
  \draw[gray!60!black, densely dotted] (0,0) -- ({\rx*cos(\xang+45)},{0.28*\rx*sin(\xang+45)});
  \draw[gray!60!black] ({0.28*\R*cos(\xang)},{0.28*0.28*\R*sin(\xang)}) arc (\xang:\xang+45:0.28*\R cm and 0.28*0.28*\R cm);
  \node[font=\scriptsize, inner sep=0pt] at ({0.7*\R*cos(\xang+24)},{0.7*0.28*\R*sin(\xang+24)-0.02}) {$\phi$};
  \foreach \k in {0,...,7} {
    \pgfmathsetmacro{\ph}{\xang+45*\k}
    \pgfmathsetmacro{\px}{\rx*cos(\ph)}
    \pgfmathsetmacro{\py}{\yc+\ry*sin(\ph)}
    \pgfmathparse{int(mod(\k,2))}
    \ifnum\pgfmathresult=0
      \fill[orange!85!black] (\px,\py) circle (0.10);
    \else
      \fill[blue!70!black] (\px,\py) circle (0.055);
    \fi
  }
  \node[blue!70!black, anchor=west, font=\footnotesize] at (\rx+0.05,\yc+0.12) {latitude $p$};
  \node[anchor=north, align=center, font=\footnotesize] at (0,-\R-0.35)
    {$K$ branches $\ket{v(\theta,\phi_\kappa)}^{\otimes L}$};
  \draw[->, very thick] (\R+0.6,0.1) -- (\R+2.0,0.1)
    node[midway, above, align=center, font=\footnotesize] {projection};
  \node[anchor=west] (D) at (\R+2.1,0.1) {\Large$\ket{D^L_n}$};
\end{tikzpicture}}
\caption{A Dicke state as the magnetization projection of a product of tilted spins (Eq.~\eqref{eq:proj}). The $K$ branches (eight shown) lie on a single latitude $p=\sin^2\frac{\theta}{2}$ of the Bloch sphere with varying azimuth $\phi$. The least magical branches (orange) lie on the $xz$ and $yz$ planes.}
\label{fig:proj}
\end{figure}

Consider a single tilted spin,
\begin{equation}\label{eq:spin}
\ket{v(\theta,\phi)}=\cos\tfrac{\theta}{2}\ket{0}+e^{i\phi}\sin\tfrac{\theta}{2}\ket{1},
\end{equation}
where $\phi$ is the azimuthal angle, and $p=\sin^2\tfrac{\theta}{2}$ is referred to as the latitude 
(Fig.~\ref{fig:proj}). A product of $L$ such spins has on average $pL$ excitations, so the latitude plays a similar role to that of the filling for a Dicke state. A discrete Fourier sum over $K\ge L+1$ product states with equally spaced azimuths $\phi_\kappa=\phi_0+2\pi \kappa/K$ selects the component with $n$ excitations (i.e.~a Dicke state),
\begin{equation}\label{eq:proj}
\frac{1}{K}\sum_{\kappa=0}^{K-1}e^{-in\phi_\kappa}\ket{v(\theta,\phi_\kappa)}^{\otimes L}=\sqrt{b(n;L,p)}\,\ket{D^L_n},
\end{equation}
where the binomial weight
\begin{equation}\label{eq:binom}
b(n;L,p)=\binom{L}{n}p^n(1-p)^{L-n}
\end{equation}
is the probability of finding $n$ excitations in $\ket{v(\theta,\phi_\kappa)}^{\otimes L}$. We call the $K$ product states $\ket{v(\theta,\phi_\kappa)}^{\otimes L}$ the ``branches'' of this superposition. 
Physically, the projection picture of Eq.~\eqref{eq:proj} is the familiar statement that a state of definite particle number is a phase average of states with definite phase, as in the number projection of a superconductor or superfluid. Under this analogy, the ``particles'' are the excitations, and the phase is the azimuth of the spins (Fig.~\ref{fig:proj}).

The branches of Eq.~\eqref{eq:proj} all lie on a single latitude $p$, while their magic depends on the azimuth. For fixed latitude, the single-qubit magic of Eq.~\eqref{eq:M2qubit} is smallest at the four azimuths $\phi\in\frac{\pi}{2}\mathbb Z$, where it equals
\begin{equation}\label{eq:acoh}
\acoh(p)=-\log_2\big[1-4p(1-p)+16p^2(1-p)^2\big].
\end{equation}
This vanishes at the poles ($p=0,1$) and the equator ($p=1/2$), since these include stabilizer states, and is maximal with the $T$-state value $\acoh(\fs)=\log_2\frac{4}{3}$, with $\fs=\frac{1}{2}(1-1/\sqrt2)$.

\subsection{Branch bound on magic}\label{subsec:branch_bound}
To relate the magic of superpositions of branches to Dicke states, we first introduce the ``branch bound'', which is proved in App.~\ref{subsecapp:branch}. For any superposition $\ket{\psi}=\sum_\kappa c_\kappa\ket{w_\kappa}$ of normalized branch states,
\begin{equation}\label{eq:branch}
M_2(\psi)\ \ge\ \min_\kappa M_2(w_\kappa)\;-\;8\log_2\frac{\|c\|_1}{\|\psi\|},
\end{equation}
where $\|\psi\|=\sqrt{\braket{\psi|\psi}}$ and $\|c\|_1=\sum_\kappa|c_\kappa|$. Hence, $\ket{\psi}$ is at least as magical as its least magical branch, up to a non-positive correction term (App.~\ref{subsecapp:branch}). Roughly speaking, the correction penalizes (i.e.~makes the bound looser) cases where the wavefunction norm is small, or the branches have large cancellations. Note that the branches do not need to be orthogonal or product states\footnote{For orthogonal branches labelled by a flag qubit, a version without the correction term was proved in Ref.~\cite{Leone2024stabilizer}.}.

Applied to the projection picture of the Dicke state (Eq.~\eqref{eq:proj}), where $\|\psi\|=b(n;L,p)^{1/2}$ and $\|c\|_1=1$, we have
\begin{equation}\label{eq:coh}
M_2(\ket{D^L_n})\ \ge\ L\acoh(p)+4\log_2b(n;L,p),
\end{equation}
since the magic of each branch is lower-bounded by $L\acoh(p)$. 
By Stirling's formula, the second term behaves at fixed filling as
\begin{equation}\label{eq:stirlingmain}
4\log_2b(n;L,p)=-4LD(f\|p)-2\log_2L+O(1),
\end{equation}
where
\begin{equation}\label{eq:D}
D(f\|p)=f\log_2\frac{f}{p}+(1-f)\log_2\frac{1-f}{1-p}
\end{equation}
is the relative entropy (or Kullback-Leibler divergence) between two Bernoulli distributions with parameters $f$ and $p$. Since $p$ can be chosen freely to maximize the bound, Eq.~\eqref{eq:coh} expresses a trade-off in the choice of the latitude. At the matched latitude $p=f$, the Dicke component of the projection has weight $b(n;L,f)\propto L^{-1/2}$ so that the correction is only $2\log_2L+O(1)$ bits, while the branch magic is $L\acoh(f)$. At other latitudes, the branches can be more magical, but the Dicke component then has an exponentially small weight, leading to the extensive correction $-4LD(f\|p)$.

From the statistical mechanics viewpoint, a product of tilted spins plays the role of a (grand) canonical ensemble for the number of excitations, while the Dicke state corresponds to the microcanonical state. $b(n;L,p)$ is the probability that the canonical ensemble has $n$ excitations, i.e., that the projection onto the Dicke state succeeds. It is only polynomially small in $L$ when $p=f$ (ensemble equivalence), and exponentially small otherwise.

\subsection{The magic density}\label{subsec:closed}

The optimized lower bound in Eq.~\eqref{eq:coh} is actually tight at the level of the density, since a matching upper bound holds. The proof of the latter further provides analytically the logarithmic and constant part of $M_2$ in the thermodynamic limit, as summarized in Result 1 and explained in Sec.~\ref{subsec:identity}:

\begin{resultbox}{Result 1 (asymptotic magic)}
At every filling $f=n/L\in(0,1)$,
\begin{equation}\label{eq:log}
M_2(\ket{D^L_n})=L\,\alpha(f)+2\log_2L+c(f)+o(1)
\end{equation}
as $L\to\infty$, where $c(f)$ is given in App.~\ref{subsecapp:den-c}, and the magic density is the optimized branch bound,
\begin{equation}\label{eq:alpha}
\alpha(f)=\max_{p\in(0,1)}\Big[\acoh(p)-4D(f\|p)\Big],
\end{equation}
with a unique maximizing latitude $p^*(f)$ given in closed form in the lemma of App.~\ref{subsecapp:den-lower}. The proof is given in Apps.~\ref{app:lower} and \ref{app:upperbounds}.
\end{resultbox}

At half-filling, the density vanishes, $\alpha(\frac{1}{2})=0$, and the constant takes the value $c(\frac{1}{2})=2\log_2\pi-7$, so that Eq.~\eqref{eq:log} becomes
\begin{equation}\label{eq:half}
M_2(\ket{D^L_{L/2}})=2\log_2\frac{\pi L}{8}-1+o(1).
\end{equation}
This half-filling result confirms the asymptotic form that Ref.~\cite{HernandezYanes2025non} obtained from a large-$L$ approximation.

\begin{figure*}[t]
\centering
\includegraphics[width=0.9\textwidth]{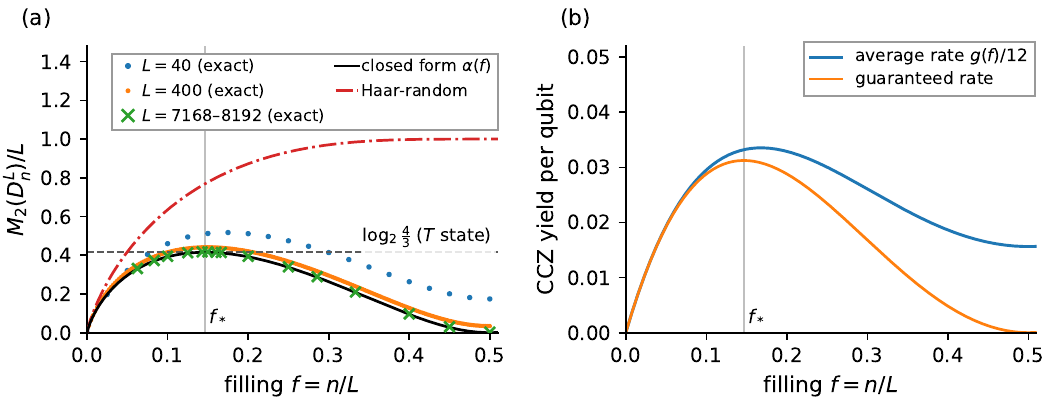}
\caption{Magic and CCZ yield of Dicke states. (a) SRE per qubit, $M_2(D^L_n)/L$. The dots show the numerically computed exact values for all $n\le L/2$ at $L=40$ (blue) and $L=400$ (orange), and the crosses show the largest computed size ($L=7168$ to $8192$) at 15 rational fillings. The black line is the closed-form magic density $\alpha(f)$ of Eq.~\eqref{eq:alpha}. At a fixed filling $f=n/L$, the difference between the exact data and $\alpha(f)$ equals  $[2\log_2L+c(f)+o(1)]/L$ (Eq.~\eqref{eq:log}). The dashed line indicates the magic of a $T$ state. The red dash-dotted curve is a lower bound for the magic density of Haar-random states at the same filling~\cite{Iannotti2026non}. 
(b) CCZ yield per qubit using the heralded eight-qubit procedure $\Lambda_8$ of Ref.~\cite{Rizzo2026universal}, applied to blocks of eight qubits. The blue curve is the average rate $g(f)/12=\frac{1}{2}f(1-f)[(1-f)^4+f^4]$, and the orange curve is the guaranteed rate $\frac{1}{2}f(1-f)(1-2f)^2$ (see Sec.~\ref{subsec:ccz}). In both panels, the vertical line denotes the filling $\fs$ where the magic density is maximal. All results are  symmetric under $f\to 1-f$.}
\label{fig:overview}
\end{figure*}

The magic density is positive at every other non-trivial filling (Fig.~\ref{fig:overview}a)\footnote{At small filling, $\alpha(f)=2f\log_2[e/(7f)]+O(f^2)$, so that $L\alpha(n/L)=2n\log_2L+O(1)$ at a fixed number $n$ of excitations. This agrees with the logarithmic magic $M_2(\ket{D^L_n})=2n\log_2L+O(1)$ at fixed $n$ (App.~\ref{subsecapp:den-lower}).}. For $0<f\le\frac{1}{2}$, its largest value, $\log_2\frac{4}{3}$ (equal to that of a $T$ state), lies below the single-qubit maximum $\log_2\frac{3}{2}$, and is reached at $f=\fs$ where the maximizing latitude $p^*$ in Eq.~\eqref{eq:alpha} equals the filling. 
Dicke states carry substantially less magic than typical states; the magic density of Haar-random states with the same filling~\cite{Iannotti2026non} is lower-bounded by the red curve in Fig.~\ref{fig:overview}a, which lies above $\alpha(f)$.

Fig.~\ref{fig:overview}a compares the analytic result with numerically exact values of $M_2$, which we have computed for every $(L,n)$ with $L\le400$, and on sequences up to $L=8192$ at various rational fillings (App.~\ref{app:numerics}).

\subsection{Walk-counting identity}\label{subsec:identity}

As detailed in App.~\ref{app:upperbounds}, Result 1 is derived by converting the summation of Eq.~\eqref{eq:M2} into a statistical mechanical problem of lattice walks. The latter involves enumerating 8-dimensional walks\footnote{The 8 dimensions can be understood as arising from the 4 replicas (due to the fourth power $\langle P\rangle^4$), times a factor of 2 from the bra and the ket.} of $L$ ``steps'', one per site, that are constrained to end at some filling-dependent point $\mathbf{n}$.
At each site, the step records the bits of the Pauli string $P$, and of the four replicas of the wavefunction. The requirement that every wavefunction configuration, in the bra and ket, has $n$ excitations fixes the endpoint $\mathbf{n}$.
This constrained ``microcanonical'' count can be evaluated by moving to a canonical ensemble of independent steps that follow a Gibbs distribution\footnote{This mapping from microcanonical to canonical ensembles parallels the discussion at the end of Sec.~\ref{subsec:branch_bound}, except here we have 8 constraints to satisfy, rather than a single constraint on the excitation number $n$.} (a related replica count has been used for a scar state of a lattice gauge theory~\cite{Ebner2026scarmagic}). Choosing the fugacities according to an effective latitude $p$ leads to the exact identity
\begin{equation}\label{eq:keymain}
\begin{aligned}
M_2(\ket{D^L_n})={}\Big[L\acoh(p)+4\log_2b(n;L,p)\Big]
-\log_2\Pr[\mathbf{n}],
\end{aligned}
\end{equation}
where the quantity in square brackets coincides with the branch bound of Eq.~\eqref{eq:coh}, and $\Pr[\mathbf{n}]$ is the probability that the weighted random walk lands at $\mathbf{n}$. This identity holds for every $L$, $n$ and latitude $p$. Since $\Pr[\mathbf n]\le1$, Eq.~\eqref{eq:keymain} recovers the branch bound, and shows that it falls short of the SRE by exactly $-\log_2\Pr[\mathbf n]$. 

Since the LHS of Eq.~\eqref{eq:keymain} does not depend on $p$, the latitude that maximizes the branch bound is the one that maximizes $\Pr[\mathbf n]$, namely the latitude $p^*$ of Eq.~\eqref{eq:alpha}, at which the average endpoint of the walk coincides with $\mathbf{n}$.
In this case, a local central limit theorem shows that $\Pr[\mathbf{n}]$ decays as $L^{-4}$ rather than exponentially, so that $-\log_2\Pr[\mathbf n]=4\log_2L+O(1)$. Combined with $4\log_2b(n;L,p^*)=-4LD(f\|p^*)-2\log_2L+O(1)$ from Eq.~\eqref{eq:stirlingmain}, this gives the logarithmic term $+2\log_2 L$ of Result 1, and implies that the magic density obeys Eq.~\eqref{eq:alpha}. The constant term $c(f)$ follows from explicit evaluation of the leading part of $\Pr[\mathbf{n}]$.

\section{Cost and yields of Dicke states}\label{sec:usefulness}

Across any bipartition, a Dicke state has at most $\min(n,L-n)+1$ Schmidt components, so that it is weakly entangled and expressible as a matrix-product state of bond dimension $\chi\leq\min(n,L-n)+1$ \cite{Raveh2024dicke} (see also App.~\ref{subsecapp:prot-marg}). Its extensive magic therefore does not signal that the state is hard to simulate classically. 
The key questions instead are how many non-Clifford resource states are needed to prepare a Dicke state (its cost), and how many can be extracted from a Dicke state (its yield). We quantify both cost and yield  in terms of the $T$ state $\ket{T}$ and the CCZ state $\ket{\CCZ}=\CCZ\ket{+}^{\otimes3}$, which can provide a $T$ gate and a Toffoli gate through gate teleportation respectively~\cite{Gottesman1999demonstrating,Zhou2000methodology}. The free operations of stabilizer quantum computation consist of Clifford unitaries, stabilizer ancillas, Pauli measurements, discarding of qubits, and classical randomness and feed-forward.

\subsection{Cost}\label{subsec:cost}

For $1\le n\le L-1$, every run that prepares $\ket{D^L_n}$ consumes at least $L-1$ $T$ states or $\lceil(L-1)/3\rceil$ CCZ states, and at half-filling at least $L-2$ $T$ states or $\lceil(L-2)/3\rceil$ CCZ states. This holds for heralded procedures, with or without a pure catalyst (App.~\ref{app:upper}). This cost lower bound arises from considering the number of Pauli strings $|\mathrm{Stab}\,\psi|$ that leave a state $\psi$ invariant up to a sign, which defines the stabilizer nullity $\nu(\psi)=L-\log_2|\mathrm{Stab}\,\psi|$. A stabilizer state has $\nu=0$, while the standard resource states have $\nu(T)=1$ and $\nu(\text{CCZ})=3$. Free operations can never increase the stabilizer nullity of a pure state, even on individual Pauli-measurement outcomes~\cite{Beverland2019lower}. The bound then follows from the fact that $\nu(D^L_n)=L-1$ (or $L-2$ at half-filling), as shown in App.~\ref{subsecapp:up-pauli}\footnote{The nullity counts Pauli symmetries rather than the spread of the Pauli spectrum, so the bound applies equally to the $W$ state ($n=1$), whose SRE grows only logarithmically~\cite{Odavic2022complexity,Catalano2024magic}.}.

A known heralded preparation of Dicke states~\cite{Yu2024efficient} involves measuring the number of excitations of $\ket{+}^{\otimes L}$, which outputs $\ket{D^L_w}$ with probability $\binom{L}{w}2^{-L}$. If the number of excitations is computed into an ancilla register~\cite{Campbell2021early}, this measurement costs $L-s_2(L)$ CCZ states, where $s_2(L)$ is the number of ones in the binary expansion of $L$ (App.~\ref{subsecapp:up-prep}). Therefore, for successful runs, the cost lower bound based on the stabilizer nullity is tight up to a factor of about 3 for CCZ states\footnote{Note that since the success probability is exponentially small in $L$ for fillings $f\neq 1/2$, targeting a specific filling typically requires repeating the procedure an exponentially large number of times.}.

\subsection{No exact $T$ states without a catalyst}\label{subsec:noT}

In the context of yields, exact $T$ states cannot be extracted from a Dicke state without a catalyst, which follows from a number-field argument of Ref.~\cite{Beverland2019lower} (App.~\ref{subsecapp:T-field}). The density matrix of a Dicke state involves only rational entries, $0$ and $\binom{L}{n}^{-1}$. When the entries of a density matrix belong to the Gaussian rationals $\QI$, they remain in $\QI$ under free operations\footnote{For example, conjugation by a Hadamard gate only introduces factors of $1/2$, and Pauli measurements only have rational probabilities.}. On the other hand, the density matrix $\ket{T}\bra{T}$ has the off-diagonal entry $e^{-i\pi/4}/2\notin\QI$. Hence, no procedure without a catalyst, even a heralded one, ever outputs an exact $T$ state from any number of Dicke states.

\subsection{CCZ yield}\label{subsec:ccz}

A CCZ state has a rational density matrix, and hence does not fall under the no-go statement of Sec.~\ref{subsec:noT}. CCZ states are therefore the natural currency\footnote{The four-qubit $W$ state $\ket{D^4_1}$ is even Clifford-equivalent to $\ket{\CCZ}\otimes\ket{0}$ (App.~\ref{subsecapp:prot-proc}).} for evaluating the yield of Dicke states (this will be converted to a $T$-state yield through catalysis in Sec.~\ref{subsec:tyield}). 
We consider heralded procedures, which announce success when the output is exact and discard failed runs otherwise. The number $t$ of CCZ states obtained from one copy of $\ket{D^L_n}$ is then a random variable. For per-qubit yields, we define the \emph{average rate} $\lim_{L\rightarrow\infty}\mathbb E[t]/L$, and the \emph{guaranteed rate}, which is the largest $r$ such that $t\ge rL$ with a probability tending to one.

Ref.~\cite{Rizzo2026universal} constructs heralded procedures that convert several copies of any unknown single-qubit state into an exact CCZ state. We  consider the eight-qubit procedure $\Lambda_8$ (App.~\ref{subsubsecapp:eight_qubit}), which succeeds on eight copies of a single-qubit state with probability $\frac{2}{3}\Mlin$. Here the linear stabilizer entropy of order three, $\Mlin(\psi)=1-2^{-L}\sum_P\langle P\rangle^6$, is another measure of the spread of the Pauli spectrum~\cite{Rizzo2026universal}. For a single qubit with Bloch vector $\mathbf{r}$, we have
\begin{equation}\label{eq:mlin}
\Mlin=\tfrac{1}{2}\big(1-r_x^6-r_y^6-r_z^6\big).
\end{equation}
We extract CCZ states from a Dicke state by acting with $\Lambda_8$ on disjoint eight-qubit blocks. While the reduced density matrix of a block of $\ket{D^L_n}$ does not correspond to a product of single-qubit states, it is close to a mixture of such products, as discussed later in Sec.~\ref{subsec:hidden}. 
This enables us to prove the following statements about the filling-dependent yield in the thermodynamic limit:
\begin{resultbox}{Result 2 (CCZ yield)}
At every filling $0<f<1$, the heralded procedure $\Lambda_8$, applied to all $N=\lfloor L/8\rfloor$ blocks of $\ket{D^L_n}$, extracts on average
\begin{equation}\label{eq:ccz}
\lim_{L\to\infty}\frac{\mathbb E[t]}{L}=\frac{1}{2}f(1-f)\big[(1-f)^4+f^4\big]
\end{equation}
CCZ states per qubit. The guaranteed rate is $\frac{1}{2}f(1-f)(1-2f)^2$. The proof is given in App.~\ref{subsecapp:prot-stats}.
\end{resultbox}
The guaranteed rate is maximized at $f_*$, and vanishes at half-filling. In contrast, the average rate is maximal at $f$ slightly above $f_*$, and remains non-vanishing at half-filling (Fig.~\ref{fig:overview}b). Interestingly, the average rate for a Dicke state coincides with that of an ensemble average over products of tilted spins $\ket{v_{f,\phi}}=\sqrt{1-f}\,\ket{0}+\sqrt{f}\,e^{i\phi}\ket{1}$ at fixed latitude $f$. Indeed, the average of $\Mlin$ over the azimuthal angle $\phi$ is
\begin{equation}\label{eq:gdef}
g(f)=\int_0^{2\pi}\frac{d\phi}{2\pi}\,\Mlin(v_{f,\phi})=6f(1-f)\big[(1-f)^4+f^4\big].
\end{equation}
Each block of such an ensemble therefore succeeds with probability $\frac{2}{3}g(f)$, which is consistent with the average rate $g(f)/12$ of Eq.~\eqref{eq:ccz}.

While typical states have a larger magic density than Dicke states (Sec.~\ref{subsec:closed}), a small block of a Haar-random state at filling $f$ is exponentially close to a mixture of computational basis states, so that procedures acting on separate blocks extract only an exponentially small number of CCZ states (App.~\ref{subsecapp:prot-marg}). 
Thus, while the magic of a Dicke state is less abundant, it is locally accessible with block procedures.

\subsection{The hidden phase}\label{subsec:hidden}

The fact that the average and guaranteed rates do not coincide in the large-$L$ limit implies that the success events of different blocks of the same Dicke state are not independent. This can be explained using the projection picture applied on any subset of $B$ qubits (which may span several blocks). As already hinted by Eq.~\eqref{eq:gdef}, the marginal $\rho_B$ of $\ket{D^L_n}$ on $B$ qubits lies within trace distance $B(B-1)/(2L)$ of an azimuthal average of tilted product states (App.~\ref{subsecapp:prot-marg}),
\begin{equation}\label{eq:phaseavg}
\rho_B\approx\int_0^{2\pi}\frac{d\phi}{2\pi}\Big(\ket{v_{f,\phi}}\bra{v_{f,\phi}}\Big)^{\otimes B},
\end{equation}
in which all $B$ qubits share the same azimuth $\phi$. The average over $\phi$ is the remnant of the projection onto a fixed magnetization, and the shared $\phi$ introduces correlations between different blocks.

\begin{figure}[t]
\centering
\includegraphics[width=1\columnwidth]{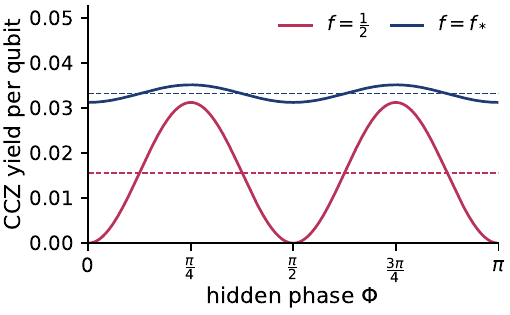}
\caption{CCZ yield per qubit $\mathcal{P}_f(\Phi)/8$ of the eight-qubit procedure $\Lambda_8$, applied to all blocks, as a function of the hidden phase $\Phi$ at $f=\frac{1}{2}$ (red) and $f=\fs$ (blue). The average yield, Eq.~\eqref{eq:ccz}, is indicated with dashed lines. The guaranteed yield is the minimum value over $\Phi$.}
\label{fig:phase}
\end{figure}

Although Eq.~\eqref{eq:phaseavg} holds only for $B\ll\sqrt L$ qubits, it determines the statistics of all $N$ blocks. Consider applying Eq.~\eqref{eq:phaseavg} to the $B=8u$ qubits of any $u$ blocks. For a given $\phi$, the $u$ blocks succeed independently, each with a probability $\mathcal{P}_f(\phi)$ corresponding to $\Lambda_8$ succeeding on eight copies of $\ket{v_{f,\phi}}$. Therefore, the probability of all $u$ blocks succeeding tends to the average of $\mathcal{P}_f(\phi)^u$ over $\phi$. Since this probability is the same for every choice of $u$ distinct blocks, it equals the $u$th moment of the success fraction $t/N$, up to corrections of order $1/N$ (App.~\ref{subsecapp:prot-stats})\footnote{This is because $(t/N)^u$ is the fraction of the $N^u$ ordered $u$-tuples of blocks in which every block succeeds, and only a fraction of order $1/N$ of these tuples contain the same block twice.}. The moments of $t/N$ therefore converge to those of $\mathcal{P}_f(\phi)$, which fixes the limiting distribution of $t/N$:
\begin{resultbox}{Result 3 (hidden phase)}
At every filling $0<f<1$, when the procedure $\Lambda_8$ is applied to all $N=\lfloor L/8\rfloor$ blocks  of $\ket{D^L_n}$, the fraction $t/N$ of successful blocks converges in distribution to the random variable
\begin{equation}\label{eq:Pphi}
\mathcal{P}_f(\Phi)=4f(1-f)\,\big|(1-f)^2-f^2e^{4i\Phi}\big|^2,
\end{equation}
with $\Phi$ uniform on $[0,2\pi)$. Here, $\mathcal{P}_f(\Phi)=\frac{2}{3}\Mlin(v_{f,\Phi})$ is the success probability of $\Lambda_8$ on eight copies of $\ket{v_{f,\Phi}}$ (Fig.~\ref{fig:phase}). The proof is given in App.~\ref{subsecapp:prot-stats}.
\end{resultbox}
Heuristically, we can think of the hidden phase $\Phi$ as a single random variable per run, shared by all blocks, and each block acts like eight copies of the same tilted spin $\ket{v_{f,\Phi}}$. 
The situation resembles that of a pair of Bose-Einstein condensates prepared with definite particle numbers. Since particle number and phase are conjugate, such a pair has no definite relative phase, yet interference experiments exhibit fringes with a definite phase that varies randomly between runs~\cite{Javanainen1996phase,Castin1997relativephase,Andrews1997observation,Hadzibabic2004interference}.

Result 3 explains several features of the CCZ yield in terms of an ensemble of tilted product states at latitude $f$ and varying azimuth $\Phi$ (Fig.~\ref{fig:phase}). 
First, the average yield is the mean of $\mathcal{P}_f(\Phi)/8$, which recovers Eq.~\eqref{eq:ccz}. Second, since $\mathcal{P}_f(\Phi)$ is not constant, two runs typically differ in their yields $t$ by $O(L)$, so that the yield exhibits macroscopic fluctuations and is not self-averaging: $t/L$ does not concentrate for any filling. The guaranteed rate is therefore the minimum $\min_\Phi\mathcal{P}_f(\Phi)/8=\frac{1}{2}f(1-f)(1-2f)^2$. Third, the guaranteed rate vanishes only at half-filling, because the equator $f=\frac{1}{2}$ passes through the four stabilizer states $\ket{\pm},\ket{\pm i}$ where $\Mlin=0$, whereas no other latitude with $0<f<1$ contains a stabilizer state.

Note that the hidden-phase property of Result 3 is not specific to $\Lambda_8$. For any fixed block procedure, the fraction of successful blocks follows the same distribution as that acting on copies of $\ket{v_{f,\Phi}}$ for large $L$ (App.~\ref{subsecapp:prot-stats}). In particular, the guaranteed rate vanishes at half-filling, since no procedure can output an exact CCZ state from the stabilizer product states $\ket{\pm}^{\otimes B}$ and $\ket{\pm i}^{\otimes B}$ on the equator. A natural question is whether learning the hidden phase from the first few blocks, and then rotating the remaining blocks toward a more favorable azimuth, improves the yield of $\Lambda_8$. This is not the case, since the only single-qubit rotations among the free operations are Clifford gates, and the success probability $\frac{2}{3}\Mlin$ of $\Lambda_8$ is unchanged when the same Clifford gate is applied to all qubits of a block.

\subsection{From CCZ to $T$ states}\label{subsec:tyield}

Since $T$ states are the most common resource state of fault-tolerant quantum computation, we also convert the CCZ yields into $T$ yields (App.~\ref{app:T}). The circuit of Ref.~\cite{Gidney2018efficient} converts one CCZ state into two $T$ states with the help of a single $T$ catalyst. A chain of such conversions, applied to the CCZ states extracted from a Dicke state using $\Lambda_8$, produces an average rate of $g(f)/6$, and a guaranteed rate of $f(1-f)(1-2f)^2$.

\subsection{Upper bounds on the yields}\label{subsec:gap}

The yields reported so far lie far below known upper bounds. Through a stabilizer nullity argument analogous to that of Sec.~\ref{subsec:cost}, we show that an exact procedure, whether or not it is heralded or uses a pure catalyst, cannot output more than one CCZ state per three qubits on any run (App.~\ref{app:upper}). 
This bound holds for any procedure, including joint and adaptive ones acting on the whole Dicke state. At half-filling, the average CCZ rate $1/64$ that we achieve by applying $\Lambda_8$ to separate blocks is a factor of $64/3\approx21$ below it.  Procedures acting on larger blocks, or jointly on many blocks, may narrow this gap, though we do not investigate the optimal rates here. Since the same nullity argument bounds the cost of preparation (Sec.~\ref{subsec:cost}), a Dicke state prepared from CCZ states never returns more CCZ states than it consumed. Its yield is therefore more relevant when the Dicke state is supplied by other means, for instance as an eigenstate of a many-body Hamiltonian (Sec.~\ref{sec:outlook}).

\subsection{Half-filling}\label{subsec:half}

We revisit half-filling, where the magic density and the guaranteed rate of block procedures vanish, while the average rate remains finite and the cost of preparation is at least linear in $L$. The cost lower bound in Sec.~\ref{subsec:cost} relies on the stabilizer nullity, which is extensive for $0<f<1$, whereas the vanishing magic density at $f=1/2$ arises from the fact that the equator contains stabilizer states in the projection picture. This contrast shows that a vanishing magic density does not necessarily indicate a poor magic resource, and points to the importance of considering different measures of magic~\cite{Liu2020many}.

The SRE does, however, bound the yield of deterministic procedures. $M_2$ cannot increase under deterministic procedures with a bounded number of steps that map pure states to each other~\cite{Leone2024stabilizer}. Since $M_2$ is additive and $M_2(\CCZ)=\log_2\frac{32}{11}$, a deterministic procedure therefore extracts at most $\alpha(f)/\log_2\frac{32}{11}$ CCZ states per qubit from a Dicke state for $L\rightarrow\infty$, and only $O(\log L)$ at half-filling. 
In contrast, $M_2$ can increase on the successful runs of heralded procedures~\cite{Haug2023stabilizer}, which explains how $\Lambda_8$ produces an extensive average yield at half-filling.

The average rate is instead more closely related to the relative entropy of magic $D_M(\psi)$~\cite{Veitch2014resource}, which measures the quantum relative entropy between $\psi$ and the closest mixture of stabilizer states. Unlike SRE, $D_M$ cannot increase on average even under heralded procedures. Result 2 therefore bounds it from below (App.~\ref{subsecapp:up-dm}),
\begin{equation}\label{eq:DMdensity}
D_M(D^L_n)\ge\log_2\frac{16}{9}\Big[\tfrac{1}{2}f(1-f)\big((1-f)^4+f^4\big)\Big]L\big(1+o(1)\big),
\end{equation}
where we used $D_M(\text{CCZ})=\log_2\frac{16}{9}$ and the additivity of $D_M$ over copies of CCZ states~\cite{Rubboli2023additivity,Rizzo2026universal}, and the filling-dependent average rate (which remains finite at half-filling) appears in square brackets.

\section{Outlook}\label{sec:outlook}

We have demonstrated that Dicke states carry extensive magic at every filling $0<f<1$ except half-filling, with a magic density obtained in closed form. We have shown that heralded block procedures extract CCZ states at a non-zero average rate per qubit, even at half-filling, while the preparation cost is extensive. The CCZ yield exhibits macroscopic fluctuations, which can be attributed to a hidden phase shared by all qubits. The magic density and the magic-state yield are intuitively understood within a projection picture, in which the Dicke state is viewed as a product of tilted spins projected onto a fixed number of excitations.

Our results can be transcribed for exact towers of eigenstates in several many-body Hamiltonians: $\eta$-pairing states in the Hubbard model~\cite{Yang1989eta,Moudgalya2020etapairing}, the many-body scar tower of the toy model of Ref.~\cite{Choi2018emergent}, and the bimagnon tower in spin-1 XY magnets~\cite{Schecter2019weak}. In suitable qubit encodings, each tower state is a Clifford image of a Dicke state, possibly with inert ancillas (App.~\ref{app:settings}). These towers therefore inherit the magic density, cost and yields of Dicke states, with the filling replaced by the pair, energy, or bimagnon density respectively. In particular, the state at the center of a scar tower has $M_2=2\log_2(\pi L/8)-1+o(1)$, while its relative entropy of magic, preparation cost and average CCZ yield remain extensive. 
The magic of quantum many-body scars has recently been studied in several models~\cite{Smith2024nonstabilizerness,Hartse2024stabilizer,Sarma2026universal,Ebner2026scarmagic}; our results show that a small SRE of a scar state does not necessarily imply that it is a weak magic resource.

It would be interesting to generalize the projection picture to other settings. Keeping $K<L+1$ branches in Eq.~\eqref{eq:proj} constrains the excitation number only modulo $K$, which raises the question of how the magic density depends on how sharply the magnetization is fixed. Another direction is to consider higher-order stabilizer R\'enyi entropies $M_k$, which would be mapped to a $4k$-dimensional lattice walk as in Sec.~\ref{subsec:identity}. Note that the inequality $M_2/(k-1)\le M_k\le M_2$ for pure states already implies that the magic is extensive for every integer $k\geq 2$, except at half-filling. Given the small bond dimension of Dicke states, it is also natural to ask if the magic is long-range, i.e., whether it survives finite-depth local circuits~\cite{Korbany2025long,Wei2025long}. Finally, it would be useful to optimize the preparation protocols and resource-state extraction procedures, to determine how tight the bounds on cost and yields are for Dicke states. The results presented here concern exact protocols, and we leave their approximate counterparts, for which the stabilizer nullity and number-field arguments no longer apply directly, to future work.

\begin{acknowledgments}
We thank Yichen Hu for helpful discussions. YK is supported by start-up funds from the University of Texas at Dallas. We acknowledge the use of AI tools Claude Opus 5.5 and Fable 5.1 for assistance with literature searches, derivations, numerical calculations, and manuscript preparation.
\end{acknowledgments}

\bibliography{refs}

@article{passarelli2024permutation,
  title = {Nonstabilizerness of permutationally invariant systems},
  author = {Passarelli, G. and Fazio, R. and Lucignano, P.},
  journal = {Phys. Rev. A},
  volume = {110},
  issue = {2},
  pages = {022436},
  numpages = {9},
  year = {2024},
  month = {Aug},
  publisher = {American Physical Society},
  doi = {10.1103/PhysRevA.110.022436},
  url = {https://link.aps.org/doi/10.1103/PhysRevA.110.022436}
}

@article{leone2022SRE,
  title = {Stabilizer {R\'enyi} Entropy},
  author = {Leone, Lorenzo and Oliviero, Salvatore F. E. and Hamma, Alioscia},
  journal = {Phys. Rev. Lett.},
  volume = {128},
  issue = {5},
  pages = {050402},
  numpages = {5},
  year = {2022},
  month = {Feb},
  publisher = {American Physical Society},
  doi = {10.1103/PhysRevLett.128.050402},
  url = {https://link.aps.org/doi/10.1103/PhysRevLett.128.050402}
}

@article{Choi2018emergent,
  title = {Emergent {SU(2)} Dynamics and Perfect Quantum Many-Body Scars},
  author = {Choi, Soonwon and Turner, Christopher J. and Pichler, Hannes and Ho, Wen Wei and Michailidis, Alexios A. and Papi{\ifmmode \acute{c}\else \'{c}\fi{}}, Zlatko and Serbyn, Maksym and Lukin, Mikhail D. and Abanin, Dmitry A.},
  journal = {Phys. Rev. Lett.},
  volume = {122},
  issue = {22},
  pages = {220603},
  numpages = {6},
  year = {2019},
  month = {Jun},
  publisher = {American Physical Society},
  doi = {10.1103/PhysRevLett.122.220603},
  url = {https://link.aps.org/doi/10.1103/PhysRevLett.122.220603}
}

@article{Schecter2019weak,
  title = {Weak Ergodicity Breaking and Quantum Many-Body Scars in Spin-1 {$XY$} Magnets},
  author = {Schecter, Michael and Iadecola, Thomas},
  journal = {Phys. Rev. Lett.},
  volume = {123},
  issue = {14},
  pages = {147201},
  numpages = {5},
  year = {2019},
  month = {Oct},
  publisher = {American Physical Society},
  doi = {10.1103/PhysRevLett.123.147201},
  url = {https://link.aps.org/doi/10.1103/PhysRevLett.123.147201}
}

@article{Moudgalya2020etapairing,
  title = {$\ensuremath{\eta}$-pairing in {Hubbard} models: From spectrum generating algebras to quantum many-body scars},
  author = {Moudgalya, Sanjay and Regnault, Nicolas and Bernevig, B. Andrei},
  journal = {Phys. Rev. B},
  volume = {102},
  issue = {8},
  pages = {085140},
  numpages = {12},
  year = {2020},
  month = {Aug},
  publisher = {American Physical Society},
  doi = {10.1103/PhysRevB.102.085140},
  url = {https://link.aps.org/doi/10.1103/PhysRevB.102.085140}
}

@inproceedings{Bartschi2019deterministic,
  title={Deterministic Preparation of {Dicke} States},
  author={B{\"a}rtschi, Andreas and Eidenbenz, Stephan},
  booktitle={Fundamentals of Computation Theory},
  pages={126--139},
  year={2019},
  organization={Springer International Publishing}
}

@article{Yu2024efficient,
  title = {Efficient Preparation of {Dicke} States},
  author = {Yu, Jeffery and Muleady, Sean R. and Wang, Yu-Xin and Schine, Nathan and Gorshkov, Alexey V. and Childs, Andrew M.},
  journal = {Phys. Rev. Lett.},
  volume = {136},
  issue = {3},
  pages = {030601},
  numpages = {9},
  year = {2026},
  month = {Jan},
  publisher = {American Physical Society},
  doi = {10.1103/9gjk-rgql},
  url = {https://link.aps.org/doi/10.1103/9gjk-rgql}
}

@misc{Joshi2026constant,
      title={Constant-Depth Unitary Preparation of {Dicke} States}, 
      author={Malvika Raj Joshi and Francisca Vasconcelos},
      year={2026},
      eprint={2601.10693},
      archivePrefix={arXiv},
      primaryClass={quant-ph},
      url={https://arxiv.org/abs/2601.10693}, 
}

@article{HernandezYanes2025non,
  title = {Nonstabilizerness in quantum-enhanced metrological protocols},
  author = {Hern{\'a}ndez-Yanes, Tanaus{\'u} and Sierant, Piotr and Zakrzewski, Jakub and P{\l{}}odzie{\ifmmode \acute{n}\else \'{n}\fi{}}, Marcin},
  journal = {Phys. Rev. A},
  volume = {113},
  issue = {1},
  pages = {012416},
  numpages = {16},
  year = {2026},
  month = {Jan},
  publisher = {American Physical Society},
  doi = {10.1103/tmf9-fyc2},
  url = {https://link.aps.org/doi/10.1103/tmf9-fyc2}
}

@article{Leone2024stabilizer,
  title = {Stabilizer entropies are monotones for magic-state resource theory},
  author = {Leone, Lorenzo and Bittel, Lennart},
  journal = {Phys. Rev. A},
  volume = {110},
  issue = {4},
  pages = {L040403},
  numpages = {6},
  year = {2024},
  month = {Oct},
  publisher = {American Physical Society},
  doi = {10.1103/PhysRevA.110.L040403},
  url = {https://link.aps.org/doi/10.1103/PhysRevA.110.L040403}
}

@article{Odavic2022complexity,
  title={Complexity of frustration: A new source of non-local non-stabilizerness},
  author={Odavi{\'c}, Jovan and Haug, Tobias and Torre, Gianpaolo and Hamma, Alioscia and Franchini, Fabio and Giampaolo, Salvatore Marco},
  journal={SciPost Physics},
  volume={15},
  number={4},
  pages={131},
  year={2023},
  publisher={Stichting SciPost}
}

@article{Catalano2024magic,
  title={Magic phase transition and non-local complexity in generalized {W} State},
  author={Catalano, Alberto Giuseppe and Odavi{\'c}, Jovan and Torre, Gianpaolo and Hamma, Alioscia and Franchini, Fabio and Giampaolo, Salvatore Marco},
  journal={SciPost Physics Core},
  volume={8},
  number={4},
  pages={078},
  year={2025},
  publisher={Stichting SciPost}
}

@article{Liu2025stabilizer,
  title = {Stabilizer {R\'enyi} entropy for translation-invariant matrix product states},
  author = {Liu, Lei-Yi-Nan and Yi, Su and Cui, Jian},
  journal = {Phys. Rev. A},
  volume = {113},
  issue = {6},
  pages = {062414},
  numpages = {8},
  year = {2026},
  month = {Jun},
  publisher = {American Physical Society},
  doi = {10.1103/7hzd-lv9k},
  url = {https://link.aps.org/doi/10.1103/7hzd-lv9k}
}

@misc{Liu2026quantifying,
      title={Quantifying Nonstabilizerness of Quantum Codes by Removing the Inert Background}, 
      author={Yuan Liu and Ke-Mi Xu},
      year={2026},
      eprint={2609.06177},
      archivePrefix={arXiv},
      primaryClass={quant-ph},
      url={https://arxiv.org/abs/2609.06177}, 
}

@misc{Liu2026quantifyingcodes,
      title={Quantifying Nonstabilizerness of Codeword-Stabilized Codes}, 
      author={Yuan Liu and Ke-Mi Xu},
      year={2026},
      eprint={2608.22017},
      archivePrefix={arXiv},
      primaryClass={quant-ph},
      url={https://arxiv.org/abs/2608.22017}, 
}

@article{Jones2012novel,
  title = {Low-overhead constructions for the fault-tolerant {Toffoli} gate},
  author = {Jones, Cody},
  journal = {Phys. Rev. A},
  volume = {87},
  issue = {2},
  pages = {022328},
  numpages = {4},
  year = {2013},
  month = {Feb},
  publisher = {American Physical Society},
  doi = {10.1103/PhysRevA.87.022328},
  url = {https://link.aps.org/doi/10.1103/PhysRevA.87.022328}
}

@article{Gidney2018efficient,
  title={Efficient magic state factories with a catalyzed {$|CCZ\rangle$} to {$2|T\rangle$} transformation},
  author={Gidney, Craig and Fowler, Austin G.},
  journal={Quantum},
  volume={3},
  pages={135},
  year={2019},
  publisher={Verein zur Forderung des Open Access Publizierens in den Quantenwissenschaften}
}

@article{Beverland2019lower,
  title={Lower bounds on the non-{Clifford} resources for quantum computations},
  author={Beverland, Michael and Campbell, Earl and Howard, Mark and Kliuchnikov, Vadym},
  journal={Quantum Science and Technology},
  volume={5},
  number={3},
  pages={035009},
  year={2020},
  publisher={IOP Publishing}
}

@misc{Rizzo2026universal,
      title={Universal magic state concentration}, 
      author={Jacopo Rizzo and Lorenzo Leone},
      year={2026},
      eprint={2608.13376},
      archivePrefix={arXiv},
      primaryClass={quant-ph},
      url={https://arxiv.org/abs/2608.13376}, 
}

@article{Rubboli2023additivity,
  title={Mixed-state additivity properties of magic monotones based on quantum relative entropies for single-qubit states and beyond},
  author={Rubboli, Roberto and Takagi, Ryuji and Tomamichel, Marco},
  journal={Quantum},
  volume={8},
  pages={1492},
  year={2024},
  publisher={Verein zur Forderung des Open Access Publizierens in den Quantenwissenschaften}
}

@article{Javanainen1996phase,
  title = {Quantum Phase of a {Bose}-{Einstein} Condensate with an Arbitrary Number of Atoms},
  author = {Javanainen, Juha and Yoo, Sung Mi},
  journal = {Phys. Rev. Lett.},
  volume = {76},
  issue = {2},
  pages = {161--164},
  numpages = {0},
  year = {1996},
  month = {Jan},
  publisher = {American Physical Society},
  doi = {10.1103/PhysRevLett.76.161},
  url = {https://link.aps.org/doi/10.1103/PhysRevLett.76.161}
}

@article{Castin1997relativephase,
  title = {Relative phase of two {Bose}-{Einstein} condensates},
  author = {Castin, Yvan and Dalibard, Jean},
  journal = {Phys. Rev. A},
  volume = {55},
  issue = {6},
  pages = {4330--4337},
  numpages = {0},
  year = {1997},
  month = {Jun},
  publisher = {American Physical Society},
  doi = {10.1103/PhysRevA.55.4330},
  url = {https://link.aps.org/doi/10.1103/PhysRevA.55.4330}
}

@article{Andrews1997observation,
  title={Observation of interference between two {Bose} condensates},
  author={Andrews, M. R. and Townsend, C. G. and Miesner, H.-J. and Durfee, D. S. and Kurn, D. M. and Ketterle, W.},
  journal={Science},
  volume={275},
  number={5300},
  pages={637--641},
  year={1997},
  publisher={American Association for the Advancement of Science (AAAS)}
}

@article{Hadzibabic2004interference,
  title = {Interference of an Array of Independent {Bose}-{Einstein} Condensates},
  author = {Hadzibabic, Zoran and Stock, Sabine and Battelier, Baptiste and Bretin, Vincent and Dalibard, Jean},
  journal = {Phys. Rev. Lett.},
  volume = {93},
  issue = {18},
  pages = {180403},
  numpages = {4},
  year = {2004},
  month = {Oct},
  publisher = {American Physical Society},
  doi = {10.1103/PhysRevLett.93.180403},
  url = {https://link.aps.org/doi/10.1103/PhysRevLett.93.180403}
}

@article{Bravyi2019simulation,
  title={Simulation of quantum circuits by low-rank stabilizer decompositions},
  author={Bravyi, Sergey and Browne, Dan and Calpin, Padraic and Campbell, Earl and Gosset, David and Howard, Mark},
  journal={Quantum},
  volume={3},
  pages={181},
  year={2019},
  publisher={Verein zur Forderung des Open Access Publizierens in den Quantenwissenschaften}
}

@article{Hartse2024stabilizer,
  title = {Stabilizer Scars},
  author = {Hartse, Jeremy and Fidkowski, Lukasz and Mueller, Niklas},
  journal = {Phys. Rev. Lett.},
  volume = {135},
  issue = {6},
  pages = {060402},
  numpages = {9},
  year = {2025},
  month = {Aug},
  publisher = {American Physical Society},
  doi = {10.1103/n5hb-l5p5},
  url = {https://link.aps.org/doi/10.1103/n5hb-l5p5}
}

@misc{Sarma2026universal,
      title={Universal equilibrium magic in quantum many-body systems}, 
      author={Soumyadeep Sarma and Tobias Haug and John Preskill and Wai-Keong Mok},
      year={2026},
      eprint={2608.22939},
      archivePrefix={arXiv},
      primaryClass={quant-ph},
      url={https://arxiv.org/abs/2608.22939}, 
}

@misc{Ebner2026scarmagic,
      title={Global and nonlocal magic of quantum many-body scars}, 
      author={Lukas Ebner and Giovanni Cataldi and Kaidi Xu and Jad C. Halimeh and Emanuele Tirrito},
      year={2026},
      eprint={2610.00129},
      archivePrefix={arXiv},
      primaryClass={quant-ph},
      url={https://arxiv.org/abs/2610.00129}, 
}

@misc{Iannotti2026non,
      title={Non-stabilizerness and {U(1)} symmetry in chaotic many-body quantum systems}, 
      author={Daniele Iannotti and Angelo Russotto and Barbara Jasser and Jovan Odavi{\'c} and Alioscia Hamma},
      year={2026},
      eprint={2603.28870},
      archivePrefix={arXiv},
      primaryClass={quant-ph},
      url={https://arxiv.org/abs/2603.28870}, 
}

@article{Dicke1954coherence,
  title = {Coherence in Spontaneous Radiation Processes},
  author = {Dicke, R. H.},
  journal = {Phys. Rev.},
  volume = {93},
  issue = {1},
  pages = {99--110},
  numpages = {0},
  year = {1954},
  month = {Jan},
  publisher = {American Physical Society},
  doi = {10.1103/PhysRev.93.99},
  url = {https://link.aps.org/doi/10.1103/PhysRev.93.99}
}

@article{Serbyn2021quantum,
  title={Quantum many-body scars and weak breaking of ergodicity},
  author={Serbyn, Maksym and Abanin, Dmitry A. and Papi{\'c}, Zlatko},
  journal={Nature Physics},
  volume={17},
  number={6},
  pages={675--685},
  year={2021},
  publisher={Springer Science and Business Media LLC}
}

@article{Moudgalya2022quantum,
  title={Quantum many-body scars and {Hilbert} space fragmentation: a review of exact results},
  author={Moudgalya, Sanjay and Bernevig, B Andrei and Regnault, Nicolas},
  journal={Reports on Progress in Physics},
  volume={85},
  number={8},
  pages={086501},
  year={2022},
  publisher={IOP Publishing}
}

@article{Chandran2023quantum,
  title={Quantum many-body scars: a quasiparticle perspective},
  author={Chandran, Anushya and Iadecola, Thomas and Khemani, Vedika and Moessner, Roderich},
  journal={Annual Review of Condensed Matter Physics},
  volume={14},
  number={1},
  pages={443--469},
  year={2023},
  publisher={Annual Reviews}
}

@article{Yang1989eta,
  title = {\ensuremath{\eta} pairing and off-diagonal long-range order in a {Hubbard} model},
  author = {Yang, Chen Ning},
  journal = {Phys. Rev. Lett.},
  volume = {63},
  issue = {19},
  pages = {2144--2147},
  numpages = {0},
  year = {1989},
  month = {Nov},
  publisher = {American Physical Society},
  doi = {10.1103/PhysRevLett.63.2144},
  url = {https://link.aps.org/doi/10.1103/PhysRevLett.63.2144}
}

@article{Kiesel2007experimental,
  title = {Experimental Observation of Four-Photon Entangled {Dicke} State with High Fidelity},
  author = {Kiesel, N. and Schmid, C. and T{\'o}th, G. and Solano, E. and Weinfurter, H.},
  journal = {Phys. Rev. Lett.},
  volume = {98},
  issue = {6},
  pages = {063604},
  numpages = {4},
  year = {2007},
  month = {Feb},
  publisher = {American Physical Society},
  doi = {10.1103/PhysRevLett.98.063604},
  url = {https://link.aps.org/doi/10.1103/PhysRevLett.98.063604}
}

@article{Prevedel2009experimental,
  title = {Experimental Realization of {Dicke} States of up to Six Qubits for Multiparty Quantum Networking},
  author = {Prevedel, R. and Cronenberg, G. and Tame, M. S. and Paternostro, M. and Walther, P. and Kim, M. S. and Zeilinger, A.},
  journal = {Phys. Rev. Lett.},
  volume = {103},
  issue = {2},
  pages = {020503},
  numpages = {4},
  year = {2009},
  month = {Jul},
  publisher = {American Physical Society},
  doi = {10.1103/PhysRevLett.103.020503},
  url = {https://link.aps.org/doi/10.1103/PhysRevLett.103.020503}
}

@article{Wieczorek2009experimental,
  title = {Experimental Entanglement of a Six-Photon Symmetric {Dicke} State},
  author = {Wieczorek, Witlef and Krischek, Roland and Kiesel, Nikolai and Michelberger, Patrick and T{\'o}th, G{\'e}za and Weinfurter, Harald},
  journal = {Phys. Rev. Lett.},
  volume = {103},
  issue = {2},
  pages = {020504},
  numpages = {4},
  year = {2009},
  month = {Jul},
  publisher = {American Physical Society},
  doi = {10.1103/PhysRevLett.103.020504},
  url = {https://link.aps.org/doi/10.1103/PhysRevLett.103.020504}
}

@article{Hume2009preparation,
  title = {Preparation of {Dicke} states in an ion chain},
  author = {Hume, D. B. and Chou, C. W. and Rosenband, T. and Wineland, D. J.},
  journal = {Phys. Rev. A},
  volume = {80},
  issue = {5},
  pages = {052302},
  numpages = {5},
  year = {2009},
  month = {Nov},
  publisher = {American Physical Society},
  doi = {10.1103/PhysRevA.80.052302},
  url = {https://link.aps.org/doi/10.1103/PhysRevA.80.052302}
}

@article{Roos2004control,
  title={Control and measurement of three-qubit entangled states},
  author={Roos, Christian F. and Riebe, Mark and H{\"a}ffner, Hartmut and H{\"a}nsel, Wolfgang and Benhelm, Jan and Lancaster, Gavin P. T. and Becher, Christoph and Schmidt-Kaler, Ferdinand and Blatt, Rainer},
  journal={Science},
  volume={304},
  number={5676},
  pages={1478--1480},
  year={2004},
  publisher={American Association for the Advancement of Science (AAAS)}
}

@article{Haffner2005scalable,
  title={Scalable multiparticle entanglement of trapped ions},
  author={H{\"a}ffner, H. and H{\"a}nsel, W. and Roos, C. F. and Benhelm, J. and Chek-al-kar, D. and Chwalla, M. and K{\"o}rber, T. and Rapol, U. D. and Riebe, M. and Schmidt, P. O. and Becher, C. and G{\"u}hne, O. and D{\"u}r, W. and Blatt, R.},
  journal={Nature},
  volume={438},
  number={7068},
  pages={643--646},
  year={2005},
  publisher={Springer Science and Business Media LLC}
}

@inproceedings{Gottesman1998heisenberg,
  title={The {Heisenberg} Representation of Quantum Computers},
  author={Gottesman, Daniel},
  booktitle={Group22: Proceedings of the XXII International Colloquium on Group Theoretical Methods in Physics},
  pages={32--43},
  year={1999},
  publisher={International Press}
}

@article{Aaronson2004improved,
  title = {Improved simulation of stabilizer circuits},
  author = {Aaronson, Scott and Gottesman, Daniel},
  journal = {Phys. Rev. A},
  volume = {70},
  issue = {5},
  pages = {052328},
  numpages = {14},
  year = {2004},
  month = {Nov},
  publisher = {American Physical Society},
  doi = {10.1103/PhysRevA.70.052328},
  url = {https://link.aps.org/doi/10.1103/PhysRevA.70.052328}
}

@article{Bravyi2005universal,
  title = {Universal quantum computation with ideal {Clifford} gates and noisy ancillas},
  author = {Bravyi, Sergey and Kitaev, Alexei},
  journal = {Phys. Rev. A},
  volume = {71},
  issue = {2},
  pages = {022316},
  numpages = {14},
  year = {2005},
  month = {Feb},
  publisher = {American Physical Society},
  doi = {10.1103/PhysRevA.71.022316},
  url = {https://link.aps.org/doi/10.1103/PhysRevA.71.022316}
}

@article{Gottesman1999demonstrating,
  title={Demonstrating the viability of universal quantum computation using teleportation and single-qubit operations},
  author={Gottesman, Daniel and Chuang, Isaac L.},
  journal={Nature},
  volume={402},
  number={6760},
  pages={390--393},
  year={1999},
  publisher={Springer Science and Business Media LLC}
}

@article{Oliviero2022magicstate,
  title = {Magic-state resource theory for the ground state of the transverse-field {Ising} model},
  author = {Oliviero, Salvatore F. E. and Leone, Lorenzo and Hamma, Alioscia},
  journal = {Phys. Rev. A},
  volume = {106},
  issue = {4},
  pages = {042426},
  numpages = {6},
  year = {2022},
  month = {Oct},
  publisher = {American Physical Society},
  doi = {10.1103/PhysRevA.106.042426},
  url = {https://link.aps.org/doi/10.1103/PhysRevA.106.042426}
}

@article{Haug2022quantifying,
  title = {Quantifying nonstabilizerness of matrix product states},
  author = {Haug, Tobias and Piroli, Lorenzo},
  journal = {Phys. Rev. B},
  volume = {107},
  issue = {3},
  pages = {035148},
  numpages = {10},
  year = {2023},
  month = {Jan},
  publisher = {American Physical Society},
  doi = {10.1103/PhysRevB.107.035148},
  url = {https://link.aps.org/doi/10.1103/PhysRevB.107.035148}
}

@article{Lami2023nonstabilizerness,
  title = {Nonstabilizerness via Perfect {Pauli} Sampling of Matrix Product States},
  author = {Lami, Guglielmo and Collura, Mario},
  journal = {Phys. Rev. Lett.},
  volume = {131},
  issue = {18},
  pages = {180401},
  numpages = {6},
  year = {2023},
  month = {Oct},
  publisher = {American Physical Society},
  doi = {10.1103/PhysRevLett.131.180401},
  url = {https://link.aps.org/doi/10.1103/PhysRevLett.131.180401}
}

@article{Tarabunga2023manybody,
  title = {Many-Body Magic Via {Pauli}-{Markov} Chains---From Criticality to Gauge Theories},
  author = {Tarabunga, Poetri Sonya and Tirrito, Emanuele and Chanda, Titas and Dalmonte, Marcello},
  journal = {PRX Quantum},
  volume = {4},
  issue = {4},
  pages = {040317},
  numpages = {19},
  year = {2023},
  month = {Oct},
  publisher = {American Physical Society},
  doi = {10.1103/PRXQuantum.4.040317},
  url = {https://link.aps.org/doi/10.1103/PRXQuantum.4.040317}
}

@article{Haug2023scalable,
  title = {Scalable Measures of Magic Resource for Quantum Computers},
  author = {Haug, Tobias and Kim, M.S.},
  journal = {PRX Quantum},
  volume = {4},
  issue = {1},
  pages = {010301},
  numpages = {23},
  year = {2023},
  month = {Jan},
  publisher = {American Physical Society},
  doi = {10.1103/PRXQuantum.4.010301},
  url = {https://link.aps.org/doi/10.1103/PRXQuantum.4.010301}
}

@article{Jordan1928uber,
  title={{{\"U}ber das Paulische {\"A}quivalenzverbot}},
  author={Jordan, P. and Wigner, E.},
  journal={Zeitschrift f{\"u}r Physik},
  volume={47},
  number={9-10},
  pages={631--651},
  year={1928},
  publisher={Springer Science and Business Media LLC}
}

@article{Korbany2025long,
  title = {Long-Range Nonstabilizerness and Phases of Matter},
  author = {Korbany, David Aram and Gullans, Michael J. and Piroli, Lorenzo},
  journal = {Phys. Rev. Lett.},
  volume = {135},
  issue = {16},
  pages = {160404},
  numpages = {7},
  year = {2025},
  month = {Oct},
  publisher = {American Physical Society},
  doi = {10.1103/1hlj-h6t9},
  url = {https://link.aps.org/doi/10.1103/1hlj-h6t9}
}

@misc{Wei2025long,
      title={Long-range nonstabilizerness and quantum codes, phases, and complexity}, 
      author={Fuchuan Wei and Zi-Wen Liu},
      year={2026},
      eprint={2503.04566},
      archivePrefix={arXiv},
      primaryClass={quant-ph},
      url={https://arxiv.org/abs/2503.04566}, 
}

@article{Gidney2017halving,
  title={Halving the cost of quantum addition},
  author={Gidney, Craig},
  journal={Quantum},
  volume={2},
  pages={74},
  year={2018},
  publisher={Verein zur Forderung des Open Access Publizierens in den Quantenwissenschaften}
}

@article{Campbell2021early,
  title={Early fault-tolerant simulations of the {Hubbard} model},
  author={Campbell, Earl T.},
  journal={Quantum Science and Technology},
  volume={7},
  number={1},
  pages={015007},
  year={2021},
  publisher={IOP Publishing}
}

@article{Haug2023stabilizer,
  title={Stabilizer entropies and nonstabilizerness monotones},
  author={Haug, Tobias and Piroli, Lorenzo},
  journal={Quantum},
  volume={7},
  pages={1092},
  year={2023},
  publisher={Verein zur Forderung des Open Access Publizierens in den Quantenwissenschaften}
}

@article{Toth2010permutationally,
  title = {Permutationally Invariant Quantum Tomography},
  author = {T{\'o}th, G. and Wieczorek, W. and Gross, D. and Krischek, R. and Schwemmer, C. and Weinfurter, H.},
  journal = {Phys. Rev. Lett.},
  volume = {105},
  issue = {25},
  pages = {250403},
  numpages = {4},
  year = {2010},
  month = {Dec},
  publisher = {American Physical Society},
  doi = {10.1103/PhysRevLett.105.250403},
  url = {https://link.aps.org/doi/10.1103/PhysRevLett.105.250403}
}

@article{Moroder2012permutationally,
  title={Permutationally invariant state reconstruction},
  author={Moroder, Tobias and Hyllus, Philipp and T{\'o}th, G{\'e}za and Schwemmer, Christian and Niggebaum, Alexander and Gaile, Stefanie and G{\"u}hne, Otfried and Weinfurter, Harald},
  journal={New Journal of Physics},
  volume={14},
  number={10},
  pages={105001},
  year={2012},
  publisher={IOP Publishing}
}

@article{Ouyang2022robust,
  title={Robust Quantum Metrology With Explicit Symmetric States},
  author={Ouyang, Yingkai and Shettell, Nathan and Markham, Damian},
  journal={IEEE Transactions on Information Theory},
  volume={68},
  number={3},
  pages={1809--1821},
  year={2022},
  publisher={Institute of Electrical and Electronics Engineers (IEEE)}
}

@article{Chen2023magic,
  title={Magic of quantum hypergraph states},
  author={Chen, Junjie and Yan, Yuxuan and Zhou, You},
  journal={Quantum},
  volume={8},
  pages={1351},
  year={2024},
  publisher={Verein zur Forderung des Open Access Publizierens in den Quantenwissenschaften}
}

@article{Robbins1955remark,
  title={A Remark on {Stirling}'s Formula},
  author={Robbins, Herbert},
  journal={The American Mathematical Monthly},
  volume={62},
  number={1},
  pages={26},
  year={1955},
  publisher={Informa UK Limited}
}

@article{Diaconis1980finite,
  title={Finite Exchangeable Sequences},
  author={Diaconis, P. and Freedman, D.},
  journal={The Annals of Probability},
  volume={8},
  number={4},
  pages={745--764},
  year={1980},
  publisher={Institute of Mathematical Statistics}
}

@article{Veitch2014resource,
  title={The resource theory of stabilizer quantum computation},
  author={Veitch, Victor and Hamed Mousavian, S A and Gottesman, Daniel and Emerson, Joseph},
  journal={New Journal of Physics},
  volume={16},
  number={1},
  pages={013009},
  year={2014},
  publisher={IOP Publishing}
}

@article{Lucke2011twin,
  title={Twin Matter Waves for Interferometry Beyond the Classical Limit},
  author={L{\"u}cke, B. and Scherer, M. and Kruse, J. and Pezz{\'e}, L. and Deuretzbacher, F. and Hyllus, P. and Topic, O. and Peise, J. and Ertmer, W. and Arlt, J. and Santos, L. and Smerzi, A. and Klempt, C.},
  journal={Science},
  volume={334},
  number={6057},
  pages={773--776},
  year={2011},
  publisher={American Association for the Advancement of Science (AAAS)}
}

@article{Zou2018beating,
  title={Beating the classical precision limit with spin-1 {Dicke} states of more than 10,000 atoms},
  author={Zou, Yi-Quan and Wu, Ling-Na and Liu, Qi and Luo, Xin-Yu and Guo, Shuai-Feng and Cao, Jia-Hao and Tey, Meng Khoon and You, Li},
  journal={Proceedings of the National Academy of Sciences},
  volume={115},
  number={25},
  pages={6381--6385},
  year={2018},
  publisher={National Academy of Sciences}
}

@article{Raveh2024dicke,
  title = {{Dicke} states as matrix product states},
  author = {Raveh, David and Nepomechie, Rafael I.},
  journal = {Phys. Rev. A},
  volume = {110},
  issue = {5},
  pages = {052438},
  numpages = {8},
  year = {2024},
  month = {Nov},
  publisher = {American Physical Society},
  doi = {10.1103/PhysRevA.110.052438},
  url = {https://link.aps.org/doi/10.1103/PhysRevA.110.052438}
}

@article{Smith2024nonstabilizerness,
  title = {Nonstabilizerness in kinetically constrained {Rydberg} atom arrays},
  author = {Smith, Ryan and Papi{\ifmmode \acute{c}\else \'{c}\fi{}}, Zlatko and Hallam, Andrew},
  journal = {Phys. Rev. B},
  volume = {111},
  issue = {24},
  pages = {245148},
  numpages = {11},
  year = {2025},
  month = {Jun},
  publisher = {American Physical Society},
  doi = {10.1103/jz4d-vdhj},
  url = {https://link.aps.org/doi/10.1103/jz4d-vdhj}
}

@misc{Turkeshi2026pauli,
      title={{Pauli} spectrum and nonstabilizerness of random fermionic {Gaussian} states}, 
      author={Xhek Turkeshi and Piotr Sierant and Poetri Sonya Tarabunga},
      year={2026},
      eprint={2609.15739},
      archivePrefix={arXiv},
      primaryClass={quant-ph},
      url={https://arxiv.org/abs/2609.15739}, 
}

@article{Zhou2000methodology,
  title = {Methodology for quantum logic gate construction},
  author = {Zhou, Xinlan and Leung, Debbie W. and Chuang, Isaac L.},
  journal = {Phys. Rev. A},
  volume = {62},
  issue = {5},
  pages = {052316},
  numpages = {12},
  year = {2000},
  month = {Oct},
  publisher = {American Physical Society},
  doi = {10.1103/PhysRevA.62.052316},
  url = {https://link.aps.org/doi/10.1103/PhysRevA.62.052316}
}

@article{Liu2020many,
  title = {Many-Body Quantum Magic},
  author = {Liu, Zi-Wen and Winter, Andreas},
  journal = {PRX Quantum},
  volume = {3},
  issue = {2},
  pages = {020333},
  numpages = {18},
  year = {2022},
  month = {May},
  publisher = {American Physical Society},
  doi = {10.1103/PRXQuantum.3.020333},
  url = {https://link.aps.org/doi/10.1103/PRXQuantum.3.020333}
}

@article{Pezze2018quantum,
  title = {Quantum metrology with nonclassical states of atomic ensembles},
  author = {Pezz{\`e}, Luca and Smerzi, Augusto and Oberthaler, Markus K. and Schmied, Roman and Treutlein, Philipp},
  journal = {Rev. Mod. Phys.},
  volume = {90},
  issue = {3},
  pages = {035005},
  numpages = {70},
  year = {2018},
  month = {Sep},
  publisher = {American Physical Society},
  doi = {10.1103/RevModPhys.90.035005},
  url = {https://link.aps.org/doi/10.1103/RevModPhys.90.035005}
}

@article{Howard2017application,
  title = {Application of a Resource Theory for Magic States to Fault-Tolerant Quantum Computing},
  author = {Howard, Mark and Campbell, Earl},
  journal = {Phys. Rev. Lett.},
  volume = {118},
  issue = {9},
  pages = {090501},
  numpages = {6},
  year = {2017},
  month = {Mar},
  publisher = {American Physical Society},
  doi = {10.1103/PhysRevLett.118.090501},
  url = {https://link.aps.org/doi/10.1103/PhysRevLett.118.090501}
}

\clearpage
\newpage
\onecolumngrid
\section*{Appendices}

The appendices provide the proofs of the results stated in the main text. By proving a branch bound, App.~\ref{app:lower} shows that the magic density of a Dicke state is at least $\alpha(f)$, which is obtained analytically. 
App.~\ref{app:upperbounds} proves the matching upper bound through an exact walk-counting formula, which completes Result 1, and derives the logarithmic correction and the constant $c(f)$. 
App.~\ref{app:numerics} describes the exact numerical evaluation of $M_2$ at large $L$ used in Figs.~\ref{fig:overview}a and \ref{fig:residual}. 
App.~\ref{app:protocols}  proves Results 2 and 3 on the CCZ  yields of Dicke states using heralded procedures. App.~\ref{app:T} shows that no exact $T$ state can be extracted without a catalyst, and converts the CCZ yield into a $T$-state yield. App.~\ref{app:upper} discusses the cost lower bound (including an explicit preparation protocol), the upper bounds on the yields, and the bound on the relative entropy of magic. App.~\ref{app:settings} identifies three examples of exact towers of states which consist of Clifford images of Dicke states.

\appendix

\tableofcontents

\clearpage\newpage
\section{Projection picture and lower bound on magic density}\label{app:lower}

Result 1 states that in the thermodynamic limit $L\rightarrow\infty$, the magic $M_2$ of a Dicke state tends to a density $\alpha(f)$ per qubit at every filling $f=n/L$. The next leading terms of $M_2$ are $2\log_2L+c(f)$, up to a remainder that vanishes as $L\to\infty$. 

In this appendix section, we use the projection picture of Sec.~\ref{subsec:mechanism} to first derive a lower bound on $M_2$. Recall that the Dicke state can be viewed as the component with exactly $n$ excitations of a product of $L$ identical tilted spins. This magnetization projection can be performed by taking a superposition of product states at fixed latitude but differing azimuth. We prove a general lower bound on the magic of superposition states (the ``branch bound''), and apply it to Dicke states. Appendix~\ref{app:upperbounds} proves the matching upper bound, and the subleading logarithmic and constant terms, from an exact walk-counting formula. Throughout, we denote the normalized sum $\mathcal{W}(\psi)$ over $4^L$ Pauli strings $P$ and the binary entropy $h_2(f)$ as
\begin{gather}\label{eq:den-W}
\mathcal W(\psi)=2^{-L}\sum_P\langle P\rangle^4=2^{-M_2(\psi)}\\
h_2(f)=-f\log_2f-(1-f)\log_2(1-f).
\end{gather}

\subsection{The branch bound}\label{subsecapp:branch}

The branch bound (Eq.~\eqref{eq:branch}) says that a superposition of states (denoted branches) is at least as magical as its least magical branch, up to corrections set by cancellations among the branches.

\medskip\noindent\emph{Lemma (branch bound).} Let $\ket{w_1},\dots,\ket{w_K}$ be $K$ normalized states of $L$ qubits, which do not need to be orthogonal, distinct or product states. Consider complex amplitudes $c_\kappa$ such that
\begin{equation}\label{eq:branch-psi}
\ket{\psi}=\sum_{\kappa=1}^Kc_\kappa\ket{w_\kappa}\neq0 ,
\end{equation}
where $\kappa$ indexes the branches. We write $\|\psi\|=\braket{\psi|\psi}^{1/2}$, $\|c\|_1=\sum_\kappa|c_\kappa|$, and $M_2(\psi)$ for the SRE of the normalized state $\ket{\hat\psi}=\ket{\psi}/\|\psi\|$. Then
\begin{equation}\label{eq:branch-stmt}
M_2(\psi)\ \ge\ \min_{\kappa}M_2(w_\kappa)-8\log_2\frac{\|c\|_1}{\|\psi\|}.
\end{equation}
By the triangle inequality, $\|\psi\|\le\|c\|_1$, so that the subtracted term is non-negative. 

\medskip\noindent\emph{Proof.} The operator $\mathcal O=\sum_{\sigma\in \{I,X,Y,Z\}}\sigma^{\otimes4}$, acting on  four copies (replicas) of a qubit, satisfies $\mathcal O^2=4\mathcal O$, so that $\mathcal O/4$ is an orthogonal projector. Hence, $\Pi=(\mathcal O/4)^{\otimes L}$ is a projector on $L$ qubits and their replicas, and satisfies
\begin{equation}\label{eq:branch-Pi}
\sum_PP^{\otimes4}=4^L\,\Pi,
\end{equation}
up to a reordering of the tensor factors. This leads to
\begin{gather}\label{eq:branch-expand}
\sum_P\bra{\hat\psi}P\ket{\hat\psi}^4=\|\psi\|^{-8}\sum_{\bm\kappa',\bm\kappa}\Big(\prod_{r=1}^4\bar c_{\kappa'_r}
c_{\kappa_r}\Big)R(\bm\kappa',\bm\kappa)\\
R(\bm\kappa',\bm\kappa)=\sum_P\prod_{r=1}^4\bra{w_{\kappa'_r}}P\ket{w_{\kappa_r}}
=4^L\bra{W_{\bm\kappa'}}\Pi\ket{W_{\bm\kappa}},
\end{gather}
where the bold indices $\bm\kappa,\bm\kappa'$ collect the branch indices of the four replicas $r=1,\dots,4$, and $\ket{W_{\bm\kappa}}=\bigotimes_{r=1}^4\ket{w_{\kappa_r}}$ are the replica states. 

We bound the RHS of Eq.~\eqref{eq:branch-expand} in three steps.
First, $\bra{W'}\Pi\ket{W}$ is the inner product of $\Pi\ket{W'}$ and $\Pi\ket{W}$, and the Cauchy-Schwarz inequality gives
\begin{equation}
|R(\bm\kappa',\bm\kappa)|\le\sqrt{G(\bm\kappa')\,G(\bm\kappa)},\qquad
G(\bm\kappa)=R(\bm\kappa,\bm\kappa)=4^L\|\Pi W_{\bm\kappa}\|^2=\sum_P\prod_{r=1}^4\bra{w_{\kappa_r}}P\ket{w_{\kappa_r}}\ \ge0.
\end{equation}
Second, the AM-GM inequality gives 
$\prod_rx_r\le\prod_r|x_r|\le\frac{1}{4}\sum_rx_r^4$, where $x_r=\bra{w_{\kappa_r}}P\ket{w_{\kappa_r}}$ is real. Summing over $P$,
\begin{equation}
G(\bm\kappa)\le\frac{1}{4}\sum_{r=1}^4\sum_P\bra{w_{\kappa_r}}P\ket{w_{\kappa_r}}^4\le\max_\kappa\sum_P
\bra{w_\kappa}P\ket{w_\kappa}^4=2^L\,2^{-\min_\kappa M_2(w_\kappa)},
\end{equation}
so that $|R(\bm\kappa',\bm\kappa)|\le2^L2^{-\min_\kappa M_2(w_\kappa)}$ for all $\bm\kappa'$ and $\bm\kappa$. 
Third, the LHS of Eq.~\eqref{eq:branch-expand} is a sum of non-negative numbers, so it is bounded by the sum of the moduli of the terms on the RHS. Since $\sum_{\bm\kappa',\bm\kappa}\prod_r|c_{\kappa'_r}||c_{\kappa_r}|=\|c\|_1^8$,
\begin{equation}
\sum_P\bra{\hat\psi}P\ket{\hat\psi}^4\le\Big(\frac{\|c\|_1}{\|\psi\|}\Big)^8\,2^L\,2^{-\min_\kappa M_2(w_\kappa)},
\end{equation}
which gives Eq.~\eqref{eq:branch-stmt}. $\square$

\subsection{The magnetization projection}\label{subsecapp:lower-proj}

This subsection applies the branch bound to the magnetization projection of Sec.~\ref{subsec:mechanism}, which gives the lower bound Eq.~\eqref{eq:coh} at any latitude. We first prove the projection identity (Eq.~\eqref{eq:proj}) itself. Expanding a product of $L$ tilted spins $\ket{v(\theta,\phi)}=\cos\frac{\theta}{2}\ket{0}+e^{i\phi}\sin\frac{\theta}{2}\ket{1}$ in the computational basis gives
\begin{equation}
\ket{v(\theta,\phi)}^{\otimes L}=\sum_x\Big(\cos\frac{\theta}{2}\Big)^{L-|x|}\Big(\sin\frac{\theta}{2}\Big)^{|x|}e^{i|x|\phi}\ket{x},
\end{equation}
where $x$ are bitstrings. For the azimuths $\phi_\kappa=\phi_0+2\pi\kappa/K$, the Fourier sum $\frac{1}{K}\sum_{\kappa=0}^{K-1}e^{i(|x|-n)\phi_\kappa}$ equals $e^{i(|x|-n)\phi_0}$ when $K$ divides $|x|-n$, and vanishes otherwise. For $K\ge L+1$, only terms with $|x|=n$ survive. With the latitude $p=\sin^2\frac{\theta}{2}$ this gives
\begin{equation}
\frac{1}{K}\sum_{\kappa=0}^{K-1}e^{-in\phi_\kappa}\ket{v(\theta,\phi_\kappa)}^{\otimes L}
=\sqrt{\tbinom{L}{n}(1-p)^{L-n}p^n}\,\ket{D^L_n}=\sqrt{b(n;L,p)}\,\ket{D^L_n},
\end{equation}
where we have defined the binomial weight
\begin{equation}
    b(n;L,p)=\tbinom{L}{n}(1-p)^{L-n}p^n.
\end{equation}

We apply the branch bound with the $K$ branches $\ket{w_\kappa}=\ket{v(\theta,\phi_\kappa)}^{\otimes L}$ and the coefficients $c_\kappa=e^{-in\phi_\kappa}/K$, which satisfy $\|c\|_1=1$. By additivity, each branch has $M_2=L\,M_2(\ket{v(\theta,\phi_\kappa)})\ge L\acoh(p)$, where $\acoh(p)=-\log_2[1-4p(1-p)+16p^2(1-p)^2]$ is the smallest magic of a single tilted spin on this latitude (see Eq.~\eqref{eq:acoh}). This yields
\begin{equation}\label{eqapp:M2_acoh}
M_2(\ket{D^L_n})\ \ge\ L\acoh(p)+4\log_2b(n;L,p).
\end{equation}

\subsection{Lower bound and closed form}\label{subsecapp:den-lower}

This subsection shows that the magic density is at least $\alpha(f)$. In particular, the branch bound gives $M_2\ge L\alpha(f)-2\log_2L-O(1)$, where $\alpha(f)$ is obtained by a maximization over $p$ which can be evaluated in closed form. Note that the logarithm $-2\log_2L$ of the branch bound lies below the true value $+2\log_2L$ (see App.~\ref{app:upperbounds}).

\subsubsection{Lower bound}

We first observe that $b(n;L,p)=\binom{L}{n}2^{-Lh_2(f)}2^{-LD(f\|p)}$, where $D(f\|p)=f\log_2\frac{f}{p}+(1-f)\log_2\frac{1-f}{1-p}$ is the relative entropy of Bernoulli distributions with parameters $f=n/L$ and $p$. Robbins' version of Stirling's formula~\cite{Robbins1955remark} gives
\begin{equation}\label{eq:stirling}
4\log_2b(n;L,p)=-4LD(f\|p)-2\log_2\big(2\pi Lf(1-f)\big)-\epsilon_L,\qquad 0<\epsilon_L<\frac{\log_2e}{3Lf(1-f)},
\end{equation}
with $\epsilon_L$ independent of $p$. Substituting into Eq.~\eqref{eqapp:M2_acoh} and maximizing over $p$ (to make the bound as tight as possible) gives
\begin{equation}\label{eq:lowerall}
M_2(\ket{D^L_n})\ \ge\ L\max_p\big[\acoh(p)-4D(f\|p)\big]-2\log_2L-2\log_2\big(2\pi f(1-f)\big)-\epsilon_L .
\end{equation}

\subsubsection{Maximization over the latitude}

The maximization over the latitude $p$ in Eq.~\eqref{eq:lowerall} can be obtained in closed form:

\medskip\noindent\emph{Lemma (optimal latitude).} Let $0<f<1$, and let $\tau=\tau(f)>0$ be the positive root of
\begin{equation}\label{eq:tau}
(1-f)\tau^2+7(1-2f)\tau-f=0,\qquad\text{i.e.}\,\,\tau(f)=\frac{\sqrt{49-192f(1-f)}-7(1-2f)}{2(1-f)} .
\end{equation}
Viewed as a function of $p\in(0,1)$, the term $\acoh(p)-4D(f\|p)$ has one stationary point, $p^*(f)=\sqrt\tau/(1+\sqrt\tau)$, which is the global maximum, and
\begin{gather}\label{eq:alphaalg}
\alpha(f)=\max_{p}\big[\acoh(p)-4D(f\|p)\big]=4h_2(f)-\mathcal H(f)\\
\mathcal H(f)=\log_2\big(1+14\tau+\tau^2\big)-2f\log_2\tau .\label{eq:mathcalH_f}
\end{gather}

\subsubsection{Properties of $\alpha(f)$}

The function $\alpha(f)$ has the following properties:
\begin{enumerate}
\item[(a)] It is symmetric, $\alpha(1-f)=\alpha(f)$.
\item[(b)] It vanishes at half-filling, $\alpha(\frac{1}{2})=0$ with $p^*(\frac{1}{2})=\frac{1}{2}$, and $\alpha(f)>0$ for all other $f\in(0,1)$.
\item[(c)] For $f\in(0,1)$, the critical points of $\alpha(f)$ are at $\fs$, $\frac{1}{2}$ and $1-\fs$. The maximal value of $\alpha(f)$ is $\log_2\frac{4}{3}$ (equal to the $T$-state value), reached for $f=\fs,1-\fs$.  At $\fs$, we have $p^*(\fs)=\fs$. 
\item[(d)] As $f\to0$,
\begin{equation}\label{eq:den-smallf}
\alpha(f)=2f\log_2\frac{e}{7f}+O(f^2),
\end{equation}
which follows by expanding Eq.~\eqref{eq:alphaalg} with $\tau=f/7+O(f^2)$ from Eq.~\eqref{eq:tau}.
\end{enumerate}

\subsubsection{Fixed number of excitations}

At a fixed number $n$ of excitations (e.g.~the $W$ state with $n=1$, and other states with a small finite number of excitations studied in Refs.~\cite{Odavic2022complexity,Catalano2024magic,Liu2025stabilizer}), the filling $n/L$ tends to zero and the magic grows only logarithmically with $L$.

To show this, we can first construct an upper bound by noticing that such a state is spread over only $\binom{L}{n}$ basis states. Retaining only the $2^L$ Pauli strings $Z^{\mathbf z}=\prod_jZ_j^{z_j}$ built from $I$ and $Z$ can only decrease the sum $\mathcal W$ of Eq.~\eqref{eq:den-W}. Since $\langle Z^{\mathbf z}\rangle=\binom{L}{n}^{-1}\sum_{|x|=n}(-1)^{\mathbf z\cdot x}$ is the Fourier transform of the uniform distribution on the bit strings with $n$ excitations, 
Parseval's identity gives $\sum_{\mathbf z}\langle Z^{\mathbf z}\rangle^2=2^L\binom{L}{n}^{-1}$. The Cauchy-Schwarz inequality then yields $\sum_{\mathbf z}\langle Z^{\mathbf z}\rangle^4\ge2^L\binom{L}{n}^{-2}$, i.e.~$\mathcal W\ge\binom{L}{n}^{-2}$, so that
\begin{equation}\label{eq:den-edge}
M_2(\ket{D^L_n})\le2\log_2\binom{L}{n}\le2Lh_2(n/L),
\end{equation}
where the second inequality is $b(n;L,n/L)=\binom{L}{n}2^{-Lh_2(n/L)}\le1$. At fixed $n$, the inequality $\binom{L}{n}\le L^n$ caps the magic at $M_2\le2n\log_2L$. Whenever $n/L\to0$, the bound $2Lh_2(n/L)$ is a vanishing fraction of $L$, so $\alpha(0)=0=\alpha(1)$.

To bound $M_2$ from below, we combine Eq.~\eqref{eq:lowerall} with the expansion in Eq.~\eqref{eq:den-smallf}, which gives $L\alpha(n/L)=2n\log_2L-2n\log_2(7n/e)+O(n^2/L)$. The term $-2\log_2L$ of Eq.~\eqref{eq:lowerall} cancels against a contribution in $-2\log_2\big(2\pi f(1-f)\big)$, leaving $-2\log_2\big[2\pi n(1-n/L)\big]$, so that
\begin{equation}
M_2(\ket{D^L_n})\ \ge\ 2n\log_2L-2n\log_2\frac{7n}{e}-2\log_2\big[2\pi n(1-n/L)\big]-\epsilon_L+O(n^2/L),
\end{equation}
in which all terms except $2n\log_2L$ stay bounded at fixed $n$. Hence, combining both bounds yields
\begin{equation}\label{eq:den-fixedn}
M_2(\ket{D^L_n})=2n\log_2L+O(1)\qquad(L\to\infty,\ n\ \text{fixed}).
\end{equation}

\clearpage\newpage
\section{Walk-counting identity}\label{app:upperbounds}

In this section, we prove the upper half of Result 1, together with the logarithmic term $+2\log_2L$ and the constant $c(f)$ in the expansion (see Eq.~\eqref{eq:log})
\begin{equation}\label{eqapp:Dicke_M2_general_form}
M_2(\ket{D^L_n})=L\alpha(f)+2\log_2L+c(f)+o(1).\end{equation} 
We use the notation of App.~\ref{app:lower}, and write a Pauli string as
\begin{equation}\label{eq:den-Pab}
P=i^{a\cdot z}X^aZ^z,\qquad X^a=X_1^{a_1}\cdots X_L^{a_L},\quad Z^z=Z_1^{z_1}\cdots Z_L^{z_L},
\end{equation}
where the bit strings $a,z\in\{0,1\}^L$ mark the positions of the letters $X$ or $Y$ and of the letters $Z$ or $Y$, respectively. Since $Y=iXZ$, the phase $i^{a\cdot z}$ counts the $Y$'s. Note that $X^aZ^z\ket{y}=(-1)^{z\cdot y}\ket{y\oplus a}$ for every bit string $y$.

App.~\ref{subsecapp:den-second} proves the walk-counting identity Eq.~\eqref{eq:keymain} and Result 1, App.~\ref{subsecapp:den-lclt} discusses the local central limit theorem in Step 5 of App.~\ref{subsecapp:den-second}, and App.~\ref{subsecapp:den-c} evaluates the constant term $c(f)$ and analyzes half-filling.

\subsection{The walk-counting formula}\label{subsecapp:den-second}

\subsubsection{Step 1: Constrained count over bit strings}

Following Ref.~\cite{Liu2026quantifyingcodes}, we first convert the sum over Pauli strings into a count of bit strings, for any state that is a uniform superposition of bit strings. For a set $\mathcal S\subseteq\{0,1\}^L$, let $\ket{\psi_{\mathcal S}}=|\mathcal S|^{-1/2}\sum_{x\in\mathcal S}\ket{x}$ be the uniform superposition of its bit strings. Then the quantity $\mathcal W=2^{-L}\sum_P\langle P\rangle^4=2^{-M_2}$ of Eq.~\eqref{eq:den-W} is
\begin{equation}\label{eq:count}
\mathcal W(\psi_{\mathcal S})=\frac{\#\big\{(a,y_1,y_2,y_3,y_4):\ a\in\{0,1\}^L,\ y_i\in\mathcal S,\ y_i\oplus a\in\mathcal S,\
y_1\oplus y_2\oplus y_3\oplus y_4=0\big\}}{|\mathcal S|^4},
\end{equation}
where the four bit strings $y_1,\dots,y_4$ are replicas.

To see this, we write $P=i^{a\cdot z}X^aZ^z$, so that $\langle P\rangle=i^{a\cdot z} \bra{\psi_{\mathcal S}}X^aZ^z\ket{\psi_{\mathcal S}}$ with
\begin{equation}
\bra{\psi_{\mathcal S}}X^aZ^z\ket{\psi_{\mathcal S}}=\frac{1}{|\mathcal S|}\sum_{y\in\mathcal S,\ y\oplus a\in\mathcal S}(-1)^{z\cdot y}.
\end{equation}
We expand the fourth power $\langle P\rangle^4=\bra{\psi_{\mathcal S}}X^aZ^z\ket{\psi_{\mathcal S}}^4$ as a sum over four replicas $y_1,\dots,y_4$. Summing the resulting sign over the $2^L$ values of $z$ gives
\begin{equation}
\sum_{z\in\{0,1\}^L}(-1)^{z\cdot(y_1\oplus y_2\oplus y_3\oplus y_4)}=\begin{cases}2^L, & y_1\oplus y_2\oplus y_3\oplus y_4=0,\\
0, & \text{otherwise}.\end{cases}
\end{equation}
Summing over $a$, with the factor $2^{-L}$ of Eq.~\eqref{eq:den-W}, gives Eq.~\eqref{eq:count}. This is the flat-state case of the dictionary of Ref.~\cite{Liu2026quantifyingcodes}.

\subsubsection{Step 2: Eight-dimensional lattice walk}

The count of Step 1 is now rewritten as the number of walks of $L$ steps in $\mathbb Z^8$ that start at the origin and end at a fixed point $\mathbf{n}$ (to be specified below). For the Dicke state, $\mathcal S$ is the set of strings of weight $n$, and $|\mathcal S|=\binom{L}{n}$. 
Write the five bit strings $a,y_1,\dots,y_4$ as the rows of a $5\times L$ array of bits. Its $j$-th \emph{column} $\sigma_j=(a_j,y_{1j},\dots,y_{4j})\in\{0,1\}^5$ collects their bits at site $j$. The parity condition $y_1\oplus y_2\oplus y_3\oplus y_4=0$ holds for each site, so the four replica bits of every column have an even sum, which leaves 16 possible columns. There are two remaining conditions that constrain the sums over sites: $|y_i|=n$ counts the ones of replica $i$, and $|y_i\oplus a|=n$ is satisfied when, on the set of sites with $a_j=1$, replica $i$ has as many ones as zeros. We therefore assign to each column $\sigma=(a,y_1,\dots,y_4)$ the integer vector
\begin{equation}\label{eq:den-step}
\mathbf d(\sigma)=\big(y_1,\dots,y_4,\ a(2y_1-1),\dots,a(2y_4-1)\big)\in\mathbb Z^8,
\end{equation}
whose first four entries are 0 or 1 and whose last four are 0 or $\pm1$. The conditions of Eq.~\eqref{eq:count} mandate that the $L$ vectors $\mathbf d(\sigma_j)$ add up, as integers, to
\begin{equation}
\mathbf n=(n,n,n,n,0,0,0,0),
\end{equation}
so that every valid configuration is a walk of $L$ steps in $\mathbb Z^8$ that ends at $\mathbf n$. Writing $\mathcal N(L,n)$ for the number of these walks, Eq.~\eqref{eq:count} and $M_2=-\log_2\mathcal W$ give
\begin{equation}\label{eq:den-M2N}
M_2(\ket{D^L_n})=4\log_2\binom{L}{n}-\log_2\mathcal N(L,n).
\end{equation}

\subsubsection{Step 3: Tilted probability distribution}

The count $\mathcal N(L,n)$ of Step 2 is microcanonical, since the $L$ steps must add up exactly to $\mathbf n$. Here, we evaluate it by passing to a canonical ensemble, in which fugacities (i.e.~effective `chemical potentials') replace these constraints and the columns (steps) are independent. For every value of the fugacities, the microcanonical count then equals the inverse probability of one constrained configuration, multiplied by the probability that the walk meets all eight constraints. The latter is only polynomially, not exponentially, small in $L$ when the mean coincides with the constraints.
A similar replica count was evaluated in Ref.~\cite{Ebner2026scarmagic} for an exact scar state of a $\mathbb{Z}_2$ lattice gauge theory, with a transfer matrix in place of independent columns and a single fugacity for the total weight of the eight strings.

We first illustrate the procedure with a one-dimensional example of a binomial coefficient. Consider a length $L$ bit string where the bits are independent and equal to one with probability $p$. Every string with $n$ ones has the same probability $p^n(1-p)^{L-n}$, so that
\begin{equation}\label{eq:den-binex}
\binom{L}{n}=p^{-n}(1-p)^{-(L-n)}\,\Pr_p[n]\qquad\text{for every } p\in(0,1),
\end{equation}
where the probability $\Pr_p[n]$ is equivalent to the binomial weight $b(n;L,p)$ of Eq.~\eqref{eq:binom}. Using Eq.~\eqref{eq:stirling}, at a fixed filling $f=n/L$, this probability is exponentially small in $L$ when $p\ne f$, while at $p=f$ (where $n$ is the mean number of excitations) the prefactor $p^{-n}(1-p)^{-(L-n)}$ equals $2^{Lh_2(f)}$ and $\Pr_f[n]$ is of order $L^{-1/2}$. Steps 3 to 5 apply the same reasoning to the walk in eight dimensions.

We therefore look for a family of probability distributions $Q$ on the 16 columns that contains one with mean step $\mathbb E[\mathbf d]=\mathbf n/L$, whose form can be motivated in four stages.
\begin{enumerate}
\item[(i)] Guided by the canonical ensemble, we take the Gibbs form $Q(\sigma)\propto e^{\boldsymbol\beta\cdot\mathbf d(\sigma)}$, with $\boldsymbol\beta\in \mathbb R^8$.
\item[(ii)] Permuting the four replicas permutes the first four and the last four coordinates of $\mathbf d$ together, and leaves $\mathbf n/L=(f,f,f,f,0,0,0,0)$ unchanged, so we may take $\boldsymbol\beta=(\beta_y,\beta_y,\beta_y,\beta_y,\beta_u, \beta_u,\beta_u,\beta_u)$ parameterized by two real numbers.
\item[(iii)] The last four coordinates $u_i=a(2y_i-1)$ must have zero mean. They vanish on the columns with $a=0$, and the six columns with $a=1$ and two replica bits equal to one have equal weight and give $u_i=+1$ and $-1$ equally often. Therefore, $\mathbb E[u_i]=0$ requires the columns $(1,0000)$ and $(1,1111)$ to have equal weight, meaning $e^{-4\beta_u}=e^{4\beta_y+4\beta_u}$, i.e., $\beta_u=-\beta_y/2$, independently of $f$.
\item[(iv)] All eight columns with $a=1$ then have the same weight $e^{2\beta_y}$, and with the single fugacity $\tau=e^{2\beta_y}>0$ we obtain the family of probability distributions
\begin{equation}\label{eq:den-Q}
Q(\sigma)=\frac{\tau^{\,\mathbf c\cdot\mathbf d(\sigma)}}{G_\tau},\qquad \mathbf c=\tfrac{1}{4}(2,2,2,2,-1,-1,-1,-1),\qquad
G_\tau=1+14\tau+\tau^2.
\end{equation}
\end{enumerate}
The weights are listed in Table~\ref{tab:den-cols}.

\begin{table}[t!]
\centering
\begin{tabular}{l@{\quad}c@{\quad}l@{\quad}c@{\quad}c}
\toprule
type of column & number & step $\mathbf d(\sigma)$ & weight $\tau^{\mathbf c\cdot\mathbf d(\sigma)}$ & contribution to $\mathbb E[y_i]$\\
\midrule
$a=0$, no replica bit equal to one & 1 & $(0000,\,0000)$ & $1$ & $0$\\
$a=0$, two replica bits equal to one & 6 & $(y,\,0000)$, e.g.\ $(1100,\,0000)$ & $\tau$ & $3\tau/G_\tau$\\
$a=0$, all four replica bits equal to one & 1 & $(1111,\,0000)$ & $\tau^2$ & $\tau^2/G_\tau$\\
$a=1$, any replica bits (with $|y|$ even) & 8 & $(y,\,u)$, e.g.\ $(1100,\,{+}{+}{-}{-})$ & $\tau$ & $4\tau/G_\tau$\\
\bottomrule
\end{tabular}
\caption{The 16 columns $\sigma=(a,y_1,\dots,y_4)$, grouped by type, with corresponding steps $\mathbf d(\sigma)=(y,u)$ of Eq.~\eqref{eq:den-step} (entries $\pm1$ of $u$ written as $\pm$), weights under $Q$, and contributions to $\mathbb E[y_i]$.}
\label{tab:den-cols}
\end{table}

We now relate the microcanonical count $\mathcal N(L,n)$ to the probability distribution $Q$. Note that, for every $\tau>0$, every walk that ends at $\mathbf n$ has the same probability,
\begin{equation}
\prod_{j=1}^LQ(\sigma_j)=\frac{\tau^{\mathbf c\cdot\mathbf n}}{G_\tau^L}=\frac{\tau^{2n}}{G_\tau^L}.
\end{equation}
Let $S_L=\sum_{j=1}^L\mathbf d(\sigma_j)$ be the random walk with independent columns $\sigma_j$ drawn from $Q$. The number $\mathcal N(L,n)$ of walks that end at $\mathbf n$ is then the inverse $G_\tau^L\tau^{-2n}$ of this probability, multiplied by the ``end-point'' probability $\Pr_\tau[S_L=\mathbf n]$ that the random walk ends at $\mathbf n$,
\begin{equation}\label{eq:den-NP}
\mathcal N(L,n)=G_\tau^L\,\tau^{-2n}\,\Pr\nolimits_\tau[S_L=\mathbf n].
\end{equation}

The identification of the fugacity with a latitude can be motivated as follows. The product state $\ket{v}^{\otimes L}$ of tilted spins $\ket{v}=\sqrt{1-p}\ket{0}+\sqrt{p}\ket{1}$ with zero azimuth gives a bit string with $k$ ones the amplitude $(1-p)^{L/2}\big(p/(1-p)\big)^{k/2}$, i.e.~a fugacity $\sqrt{p/(1-p)}$ per excitation. If the Dicke state in Step 1 is replaced by $\ket{v}^{\otimes L}$, each column carries the product of the amplitudes of eight strings (the four replicas $y_i$ and their shifts $y_i\oplus a$). These contain $4\,\mathbf c\cdot\mathbf d(\sigma)$ ones in total (Table~\ref{tab:den-cols}), so the weight of a column is $(1-p)^4\tau^{\mathbf c\cdot\mathbf d(\sigma)}$ with $\tau=\big(p/(1-p)\big)^2$, and the 16 weights add up to $(1-p)^4G_\tau=2^{-\acoh(p)}$. The distribution $Q$ therefore describes how the least magical branch at latitude $p$ distributes $\sum_P\langle P\rangle^4$ over the columns. With this $\tau$, Eqs.~\eqref{eq:den-M2N} and \eqref{eq:den-NP} give
\begin{equation}\label{eq:key-p}
M_2(\ket{D^L_n})=L\acoh(p)+4\log_2b(n;L,p)-\log_2\Pr\nolimits_{\tau}[S_L=\mathbf n]
\end{equation}
at every latitude $p$, which is the identity Eq.~\eqref{eq:keymain} of the main text. The branch bound at latitude $p$ is this identity without the last term. The message of Eq.~\eqref{eq:key-p} is then that the Dicke state shares the same Pauli fourth moment of this branch, but restricted to the configurations in which all eight strings carry $n$ excitations.

As in the binomial example, the end-point probability is exponentially small in $L$ unless the mean step equals $\mathbf n/L$. From Table~\ref{tab:den-cols}, we can compute
\begin{equation}
\mathbb E[y_i]=\frac{7\tau+\tau^2}{G_\tau}.
\end{equation}
Multiplying by $G_\tau$, the condition $\mathbb E[y_i]=f$ becomes the quadratic $(1-f)\tau^2+(7-14f)\tau-f=0$, i.e.~Eq.~\eqref{eq:tau}, whose positive root $\tau(f)=\big(p^*/(1-p^*)\big)^2$ corresponds to the optimal latitude $p^*$ of Eq.~\eqref{eq:alpha} (lemma of App.~\ref{subsecapp:den-lower}). From now on, we denote $\tau=\tau(f)$ with $f=n/L$, so the random walk has mean $\mathbf{n}$, and write $\Pr[S_L=\mathbf n]$ for $\Pr_{\tau(f)}[S_L=\mathbf n]$. The factor $G_\tau^L\tau^{-2n}$ of Eq.~\eqref{eq:den-NP} then equals $2^{L\mathcal H(f)}$, where
\begin{equation}\label{eq:den-H}
\mathcal H(f)=\log_2G_\tau-2f\log_2\tau=-\sum_\sigma Q(\sigma)\log_2Q(\sigma)
\end{equation}
is the function of Eq.~\eqref{eq:mathcalH_f}. The second equality identifies $\mathcal H(f)$ with the entropy of the step distribution $Q$, which holds because $\log_2Q(\sigma)=(\mathbf c\cdot\mathbf d(\sigma))\log_2\tau-\log_2G_\tau$ and $\mathbf c\cdot\mathbb E[\mathbf d]=2f$. At $p=p^*$, Eq.~\eqref{eq:key-p} therefore takes the form
\begin{equation}\label{eq:key}
M_2(\ket{D^L_n})=4\log_2\binom{L}{n}-L\mathcal H(f)-\log_2\Pr[S_L=\mathbf n],
\end{equation}
and Step 5 shows that $\Pr[S_L=\mathbf n]$ is only polynomially small in $L$.

\subsubsection{Step 4: Lower bound}

Eq.~\eqref{eq:key} is exact, though at this point we know only that $0<\Pr[S_L=\mathbf n]\le1$. The probability is non-zero because the walk of $n$ columns $(0,1111)$ and $L-n$ columns $(0,0000)$ ends at $\mathbf n$. The inequality $\Pr[S_L=\mathbf n]\le1$ makes the last term of Eq.~\eqref{eq:key} non-negative, leading to the branch bound at $p^*$, i.e.~the lower bound $M_2\ge L\alpha(f)-2\log_2L-O(1)$ of Eq.~\eqref{eq:lowerall}. The matching upper bound on the magic density requires more information, namely that the end-point probability $\Pr[S_L=\mathbf n]$ is not exponentially small in $L$.

\subsubsection{Step 5: Local central limit theorem}

This step computes the end-point probability $\Pr[S_L=\mathbf n]$ to leading order, which is proportional to $L^{-4}$. This gives the term $+2\log_2L$ and the constant $c(f)$ of the Dicke state magic (see Eq.~\eqref{eqapp:Dicke_M2_general_form}).

\medskip\noindent\emph{Theorem (local limit).} Let the number of excitations $n=n_L$ grow with $L$ such that $n_L/L\to f\in(0,1)$, where $L$ may run over any (infinite) set of sizes (e.g.~the multiples of the denominator of a rational $f$, with $n_L=fL$). Let $\Sigma$ be the covariance matrix of one step under the probability distribution $Q$ at $\tau=\tau(n_L/L)$. Then
\begin{equation}\label{eq:lclt}
\Pr[S_L=\mathbf n]=\frac{32}{(2\pi L)^4\sqrt{\det\Sigma}}\big(1+o(1)\big)\qquad(L\to\infty).
\end{equation}

\medskip The proof is deferred to App.~\ref{subsecapp:den-lclt}. The $8\times8$ matrix $\Sigma$ measures the spread of one step around its mean. For establishing the magic density and the logarithmic correction, only the positivity $\det\Sigma>0$ matters, which is shown in App.~\ref{subsecapp:den-lclt}. 
The actual value of $\det\Sigma$ enters only the constant $c(f)$ (App.~\ref{subsecapp:den-c}). The choice of $\tau$ in Step 3 is crucial here; since $\mathbf n$ is the mean of $S_L$, the Gaussian that approximates the distribution of $S_L$ is evaluated at its peak, such that $\Pr[S_L=\mathbf n]$ is only polynomially small in $L$.

\subsubsection{Step 6: Magic density and the logarithmic term}

This step inserts the end-point probability $\Pr[S_L=\mathbf n]$ into the exact identity of Eq.~\eqref{eq:key} and collects the powers of $L$. Eq.~\eqref{eq:stirling} at $p=f$, where $b(n;L,f)=\binom{L}{n}2^{-Lh_2(f)}$, reads $4\log_2\binom{L}{n}=4Lh_2(f)-2\log_2(2\pi Lf(1-f))-\epsilon_L$ with $0<\epsilon_L<\log_2e/[3Lf(1-f)]$. 
Note that $\epsilon_L\to0$ because $Lf(1-f)=n(L-n)/L\to\infty$. Inserting this form and Eq.~\eqref{eq:lclt} into Eq.~\eqref{eq:key}, with $\alpha(f)=4h_2(f)-\mathcal H(f)$, gives
\begin{equation}\label{eq:den-log}
M_2(\ket{D^L_n})=L\alpha(f)+2\log_2L+\Big[2\log_2(2\pi)-2\log_2\big(f(1-f)\big)-5+\tfrac{1}{2}\log_2\det\Sigma\Big]+o(1).
\end{equation}
The bracket is evaluated in closed form in App.~\ref{subsecapp:den-c}.

The coefficient $+2$ of the logarithm can be motivated by counting powers of $L$. A walk whose step has a nondegenerate covariance spreads over a distance of order $\sqrt L$ in each direction, and hence over a volume of order $L^4$ in eight dimensions, so the probability that it ends at its mean is of order $L^{-4}$. The term $-\log_2\Pr[S_L=\mathbf n]$ therefore adds $4\log_2L$ to $M_2$. The normalization $\binom{L}{n}^4$ imposes one constraint, $|y_i|=n$, on each of the four replicas, and as illustrated by the binomial example of Eq.~\eqref{eq:den-binex}, each constraint costs a factor of order $L^{-1/2}$. Therefore, the term $4\log_2\binom{L}{n}$ contributes $-2\log_2L$, and the total coefficient of the logarithm is $+2$.

\subsubsection{Comparison with typical states}
The magic density of a Dicke state takes the form $\alpha(f)=4h_2(f)-\min_{\tau>0}\big[\log_2G_\tau-2f\log_2\tau\big]$, with $G_\tau$ defined in Eq.~\eqref{eq:den-Q}. The annealed magic density $m(f)$ of Haar-random states with the same filling~\cite{Iannotti2026non}, given there as a saddle-point expression, can be rewritten in the same form, with $G_\tau$ replaced by $1+6\tau+\tau^2$. (The same function is the annealed magic density of random free-fermion (Slater determinant) states at filling $f$~\cite{Turkeshi2026pauli}.) Since $1+14\tau+\tau^2>1+6\tau+\tau^2$, it follows that $\alpha(f)<m(f)$ for every $0<f<1$.

\subsection{Proof of the local central limit theorem}\label{subsecapp:den-lclt}

The idea of the proof of the local central limit theorem in Step 5 is to write $\Pr[S_L=\mathbf n]$ as a Fourier integral of the $L$-th power of the characteristic function of one step, whose modulus equals one at 32 equivalent wave vectors, and then evaluate a Gaussian integral around these wave vectors. The factor 32 reflects the fact that the steps of the 8-dimensional walk generate a sublattice of $\mathbb Z^8$ that contains one integer point out of every 32 points.

As in the theorem, $Q$ is the tilted probability distribution at $\tau=\tau(n_L/L)$, so that the mean step is $\mathbf n/L$, and $\Sigma$ is its covariance matrix. Strictly speaking, $Q$ generally changes slightly with $L$, but this is harmless because $\tau(n_L/L)$ converges to $\tau(f)$, the steps are bounded, and $Q$ and $\Sigma$ depend continuously on $\tau$. All bounds below hold uniformly for $\tau$ in a closed interval around $\tau(f)$, and hence for all large $L$, so we do not explicitly track the dependence on $L$.

\subsubsection{Fourier representation}

Let
\begin{equation}
\varphi(\boldsymbol\omega)=\sum_\sigma Q(\sigma)\,e^{i\boldsymbol\omega\cdot\mathbf d(\sigma)},\qquad
\boldsymbol\omega\in\mathbb R^8,
\end{equation}
be the characteristic function of one step, where $\boldsymbol\omega$ is a wave vector. Since the $L$ steps are independent, the Fourier transform of the distribution of $S_L$ is $\varphi(\boldsymbol\omega)^L$. Since the walk takes integer values, the inverse transformation on $\mathbb Z^8$ gives
\begin{equation}\label{eq:den-fourier}
\Pr[S_L=\mathbf n]=\frac{1}{(2\pi)^8}\int_{[-\pi,\pi]^8}\varphi(\boldsymbol\omega)^L\,e^{-i\boldsymbol\omega\cdot\mathbf n}\,
d\boldsymbol\omega,
\end{equation}
with the $\boldsymbol\omega$ integral taken over the cube $[-\pi,\pi]^8$.

\subsubsection{The peaks}

As every column has positive weight, the condition $|\varphi(\boldsymbol\omega)|=1$ requires the 16 phases $e^{i\boldsymbol\omega\cdot\mathbf d(\sigma)}$ to be equal. The column $(0,0000)$ with step $\mathbf d=0$ then fixes each phase factor to be 1, so that $\boldsymbol\omega\cdot\mathbf d(\sigma)$ needs to be a multiple of $2\pi$ for every column $\sigma$. We call these wave vectors the peaks. Modulo $2\pi$ in each component, they are
\begin{equation}\label{eq:den-peaks}
\begin{gathered}
\boldsymbol\omega^*=(\omega_0,\omega_0,\omega_0,\omega_0,\;\omega_0'+\pi s_1,\dots,\omega_0'+\pi s_4),\\
\omega_0\in\{0,\pi\},\quad \omega_0'\in\{0,\tfrac{\pi}{2}\},\quad s\in\{0,1\}^4,\quad s_1+\dots+s_4\ \text{even}.
\end{gathered}
\end{equation}
Note that the first four components are equal, and the last four are equal up to shifts of an even number of them by $\pi$. There are two choices of $\omega_0$, two choices of $\omega_0'$ and eight shift patterns $s$, i.e.~32 peaks in the hypercube $[-\pi,\pi]^8$. These peaks can be deduced from the conditions summarized in Table~\ref{tab:den-peaks}.

\begin{table}[t!]
\centering

\begin{tabular}{l@{\quad}l@{\quad}l}
\toprule
column & phase $\boldsymbol\omega\cdot\mathbf d(\sigma)$ & condition\\
\midrule
$a=0$, bits $i,j$ & $\omega_i+\omega_j$ & $\omega_i+\omega_j\in2\pi\mathbb Z$\\
$a=1$, no bits & $-\Omega$ & $\Omega\in2\pi\mathbb Z$\\
$a=1$, bits $i,j$ & $\omega_i+\omega_j+2(\omega_{4+i}+\omega_{4+j})-\Omega$ & $\omega_{4+i}+\omega_{4+j}\in\pi\mathbb Z$\\
other columns & combinations of the above & none new\\
\bottomrule
\end{tabular}
\caption{The conditions that each type of column of Table~\ref{tab:den-cols} imposes on a peak $\boldsymbol\omega$ such that $\boldsymbol\omega\cdot\mathbf d(\sigma)$ is a multiple of $2\pi$. Here $i\ne j$ denote the replica bits equal to one, and $\Omega=\omega_5+\omega_6+\omega_7+\omega_8$.}
\label{tab:den-peaks}
\end{table}

The integrand of Eq.~\eqref{eq:den-fourier} is unchanged by a shift by a peak wave vector $\boldsymbol\omega^*$ (since $e^{i\boldsymbol\omega^* \cdot\mathbf{d}(\sigma)}=1$), so the integral over the cube is 32 times the integral over the ``Wigner-Seitz cell'' $\mathcal V$ of wave vectors that are closer to the origin than to any other peak,
\begin{equation}\label{eq:den-fourierV}
\Pr[S_L=\mathbf n]=\frac{32}{(2\pi)^8}\int_{\mathcal V}\varphi(\boldsymbol\omega)^L\,e^{-i\boldsymbol\omega\cdot\mathbf n}\,
d\boldsymbol\omega.
\end{equation}
The origin is the only peak lying  in $\mathcal V$.

\subsubsection{Gaussian approximation}

Near the origin, we write the integrand of Eq.~\eqref{eq:den-fourierV} as $\widehat\varphi(\boldsymbol\omega)^L$, with
\begin{equation}
\widehat\varphi(\boldsymbol\omega)=e^{-i\boldsymbol\omega\cdot\mathbf n/L}\,\varphi(\boldsymbol\omega)
=\sum_\sigma Q(\sigma)\,e^{i\boldsymbol\omega\cdot(\mathbf d(\sigma)-\mathbf n/L)},
\end{equation}
which can be converted into a Gaussian form. To see this, we first note that since the mean step is $\mathbf n/L$, the term linear in $\boldsymbol\omega$ vanishes,
\begin{equation}\label{eq:den-phihat}
\widehat\varphi(\boldsymbol\omega)=1-\tfrac{1}{2}\boldsymbol\omega^{\mathsf T}\Sigma\boldsymbol\omega+O(|\boldsymbol\omega|^3).
\end{equation}
The cubic remainder is bounded by $C|\boldsymbol\omega|^3$ with a constant $C$ independent of $L$, because every entry of the centred step lies in $[-1,1]$. 
Moreover, $\Sigma$ is positive definite: $\boldsymbol\omega^{\mathsf T}\Sigma\boldsymbol\omega$ is the variance of $\boldsymbol\omega\cdot\mathbf d$, and vanishes only if $\boldsymbol\omega\cdot\mathbf d(\sigma)$ takes the same value on all 16 columns (all of which have non-zero probability in $Q$). The column $(0,0000)$ has $\mathbf d=0$, so this value is zero. The columns with $a=0$ and two replica bits $i\ne j$ then give $\omega_i+\omega_j=0$, hence $\omega_1=\dots=\omega_4=0$. The columns with $a=1$ give $\omega_5=\dots=\omega_8=0$ in the same way, so that $\boldsymbol\omega=0$.

Hence $\widehat\varphi^L\approx e^{-L\boldsymbol\omega^{\mathsf T}\Sigma\boldsymbol\omega/2}$ is a Gaussian of width of order $L^{-1/2}$. Its integral comes from $|\boldsymbol\omega|\lesssim L^{-1/2}$, and there the cubic remainder, which adds $O(L|\boldsymbol\omega|^3)=O(L^{-1/2})$ to the exponent, is negligible. The rest of the argument follows the standard proof of the local central limit theorem. On a small ball of radius $\delta$ around the origin, $|\widehat\varphi|$ is bounded by an $L$-independent Gaussian, and dominated convergence after the substitution $\boldsymbol\omega=\boldsymbol\omega'/\sqrt L$ gives
\begin{equation}
\int_{|\boldsymbol\omega|\le\delta}\widehat\varphi(\boldsymbol\omega)^L\,d\boldsymbol\omega
=\frac{(2\pi)^4}{L^4\sqrt{\det\Sigma}}\big(1+o(1)\big).
\end{equation}
On the rest of $\mathcal V$, which contains no peak, $|\varphi|\le1-\eta$ for some $\eta>0$, so that part contributes $O((1-\eta)^L)$, i.e.~$o(L^{-4})$. With the factor $32/(2\pi)^8$ of Eq.~\eqref{eq:den-fourierV}, this gives Eq.~\eqref{eq:lclt}.
\subsection{The constant term and half-filling}\label{subsecapp:den-c}

The constant term of Eq.~\eqref{eq:den-log} depends on the tilted distribution $Q$ through $\det\Sigma$.  With $\tau=\tau(f)$ (Eq.~\eqref{eq:tau}) and $G=G_\tau$ (Eq.~\eqref{eq:den-Q}), the covariance matrix and its determinant are (using Table~\ref{tab:den-cols})
\begin{equation}\label{eq:den-det}
\Sigma=\frac{\tau}{G}\begin{pmatrix}4I+\gamma J&4I\\4I&8I\end{pmatrix},\qquad\gamma=\frac{3(1-\tau)^2}{G},\qquad
\det\Sigma=\frac{2^{16}\,\tau^8\,(7\tau^2+2\tau+7)}{G^9},
\end{equation}
where $I$ is the $4\times4$ identity matrix, and $J$ is the $4\times4$ matrix with all entries equal to one. Substituting the determinant into the bracket of Eq.~\eqref{eq:den-log} gives the constant of Result 1,
\begin{equation}\label{eq:cf}
c(f)=2\log_2(2\pi)+3+4\log_2\tau+\tfrac{1}{2}\log_2\big(7\tau^2+2\tau+7\big)-\tfrac{9}{2}\log_2\big(1+14\tau+\tau^2\big)
-2\log_2\big[f(1-f)\big].
\end{equation}

At $f=\frac{1}{2}$, Eq.~\eqref{eq:tau} gives $\tau=1$, leading to the half-filling result $M_2(\ket{D^L_{L/2}})=2\log_2(\pi L/8)-1+o(1)$ of Eq.~\eqref{eq:half} at $n=L/2$.

\clearpage
\newpage

\section{Numerical computation of $M_2$}\label{app:numerics}

This appendix describes how $M_2$ is computed numerically. As for every permutation-symmetric state, $\bra{D^L_n}P\ket{D^L_n}$ depends only on the numbers $w_X,w_Y,w_Z,w_I$ of letters $X,Y,Z,I$ in $P$, and not on their positions~\cite{Toth2010permutationally,Moroder2012permutationally,Ouyang2022robust,Chen2023magic}. The sum over Pauli strings therefore reduces to a sum over these Pauli types, as in the SRE of permutation-invariant systems in Ref.~\cite{passarelli2024permutation}. For Dicke states, the value for a given type is a product of Krawtchouk polynomials~\cite{Ouyang2022robust}. 
With the notation of Eq.~\eqref{eq:den-Pab}, the string $P$ maps a basis state $y$ of weight $n$ to $y\oplus a$, which has weight $n$ only if $y$ has as many excitations as empty sites on the support of $a$, i.e., on the $w_X+w_Y=2h$ positions of the letters $X$ and $Y$. 
The $n$ excitations of such a $y$ therefore split into $h$ on the $X,Y$ positions (each of which contributes a sign when it sits on a $Y$) and $n-h$ on the $Z,I$ positions (each of which contributes a sign when it sits on a $Z$). Summing over $y$ thus gives a product of two Krawtchouk factors in
\begin{equation}\label{eq:type}
\bra{D^L_n}P\ket{D^L_n}=\frac{i^{w_Y}}{\binom{L}{n}}\,\mathcal K_h(w_Y,2h)\,\mathcal K_{n-h}(w_Z,w_Z+w_I),
\qquad \mathcal K_w(u,v)=\sum_{j}(-1)^j\binom{u}{j}\binom{v-u}{w-j}.
\end{equation}
The expectation value vanishes when $w_X+w_Y$ is odd. Here $\mathcal K_w(u,v)$ is the coefficient of $s^w$ in $(1-s)^u(1+s)^{v-u}$, a binary Krawtchouk polynomial. 
A Pauli type with $w_X+w_Y=2h$ contains $\binom{L}{2h}\binom{2h}{w_Y}\binom{L-2h}{w_Z}$ different strings, and the sum of the fourth powers over Pauli strings factorizes as
\begin{equation}\label{eq:num-sum}
\binom{L}{n}^4\sum_P\bra{D^L_n}P\ket{D^L_n}^4=\sum_{h=0}^{\min(n,L-n)}\binom{L}{2h}
\Big[\sum_{w_Y=0}^{2h}\binom{2h}{w_Y}\,\mathcal K_h(w_Y,2h)^4\Big]
\Big[\sum_{w_Z=0}^{L-2h}\binom{L-2h}{w_Z}\,\mathcal K_{n-h}(w_Z,L-2h)^4\Big].
\end{equation}

\begin{figure}[h]
\centering
\includegraphics[width=0.5\textwidth]{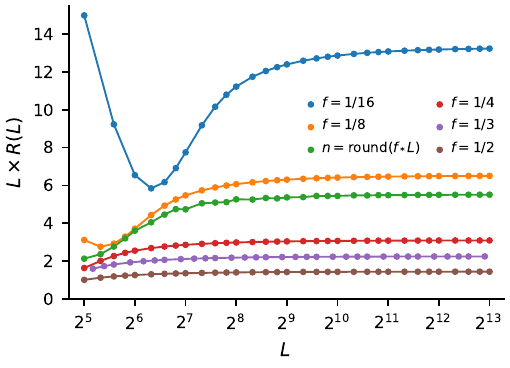}
\caption{Remainder of Result 1, $R(L)=M_2-[L\alpha(f)+2\log_2L+c(f)]$, multiplied by $L$, for Dicke states at a fixed filling $f=n/L$ (and at $n=\mathrm{round}(\fs L)$ near the peak filling), from exact values of $M_2$ up to $L=8192$.}
\label{fig:residual}
\end{figure}

$M_2$ is computed using the above for every $(L,n)$ with $2\le L\le400$, and along sequences at fixed filling up to $L=8192$ at 15 rational fillings between $\frac{1}{16}$ and $\frac{1}{2}$. We also considered sequences $n=\mathrm{round}(\fs L)$ near the filling corresponding to maximal magic density. Fig.~\ref{fig:residual} shows the remainder $R(L)=M_2-[L\alpha(f)+2\log_2L+c(f)]$ (multiplied by $L$) of the expansion of Result 1 (Eq.~\eqref{eq:log}) at several fillings. Result 1 proves that $R(L)$ vanishes as $L\to\infty$, which is consistent with the numerical data.

\clearpage
\newpage

\section{Block marginals and CCZ yield}\label{app:protocols}

In this section, we prove Results 2 and 3 on the CCZ yield of Sec.~\ref{subsec:ccz}. Both concern the heralded procedure $\Lambda_8$ of Ref.~\cite{Rizzo2026universal}, applied to the $N=\lfloor L/8\rfloor$ blocks of eight consecutive qubits of a Dicke state at filling $f=n/L$. Result~2 states that this procedure extracts on average $g(f)/12$ exact CCZ states per qubit as $L\to\infty$, where $g(f)$ is the azimuthal average of the linear stabilizer entropy of a tilted spin at latitude $f$. Result 2 also gives a guaranteed rate. Result 3 states that the fraction of successful blocks converges in distribution to the success probability $\mathcal {P}_f(\Phi)$ of $\Lambda_8$ applied to eight copies of a tilted spin whose azimuth $\Phi$ is a hidden phase. Since $\mathcal{P}_f(\Phi)$ is not constant, the yield does not concentrate for large systems.

We first derive the block marginals of a Dicke state, on which the hidden-phase picture of Sec.~\ref{subsec:hidden} rests. We then describe the procedures $\Lambda_6$ and $\Lambda_8$ of Ref.~\cite{Rizzo2026universal} together with their success probabilities, and finally derive the statistics of the blockwise yield.

\subsection{Block marginals}\label{subsecapp:prot-marg}

The yields of block procedures depend on the state only through its marginals on sets of blocks. We first determine these marginals for the Dicke state, and the distance to the density matrix $\bar\rho_B(f)$ obtained by averaging over the azimuth $\phi$ of the product states $(\ket{v_{f,\phi}}\bra{v_{f,\phi}})^{\otimes B}$ of the tilted spin $\ket{v_{f,\phi}}=\sqrt{1-f}\,\ket{0}+\sqrt f\,e^{i\phi}\ket{1}$,
\begin{equation}\label{eq:prot-rhobar}
\bar\rho_B(f)=\int_0^{2\pi}\frac{d\phi}{2\pi}\Big(\ket{v_{f,\phi}}\bra{v_{f,\phi}}\Big)^{\otimes B}.
\end{equation}
We write $\mathrm{Sym}^B$ for the symmetric subspace of $B$ qubits, spanned by $\ket{D^B_0},\dots,\ket{D^B_B}$ (i.e.~Dicke states on $B$ qubits), and $d_{\rm tr}(\rho,\sigma)=\frac{1}{2} \|\rho-\sigma\|_1$ for the trace distance between two density matrices.

\medskip\noindent\emph{Lemma D1 (block marginals).} Let $\rho_B$ be the marginal of $\ket{D^L_n}$ on any $B$ qubits.
\begin{enumerate}\renewcommand{\labelenumi}{(\roman{enumi})}
\item With the hypergeometric weights $w_j=\binom{B}{j}\binom{L-B}{n-j}/\binom{L}{n}$,
\begin{equation}\label{eq:prot-marg}
\rho_B=\sum_{j=0}^Bw_j\ket{D^B_j}\bra{D^B_j}.
\end{equation}
\emph{Proof.} Splitting each bit string $x$ of weight $n$ into its part $y$ on the block and its part $z$ on the remaining $L-B$ qubits, and grouping the strings by the weight $j$ of $y$, gives
\begin{equation}\label{eq:prot-schmidt}
\ket{D^L_n}=\sum_{j=0}^B\sqrt{w_j}\,\ket{D^B_j}\ket{D^{L-B}_{n-j}},
\end{equation}
because the number of strings with $|y|=j$ is $\binom{B}{j}\binom{L-B}{n-j}$. Eq.~\eqref{eq:prot-schmidt} is a Schmidt decomposition, and tracing out the second factor gives Eq.~\eqref{eq:prot-marg}. Note that as a result, across any bipartition, a Dicke state has at most $\min(n,L-n)+1$ Schmidt components, meaning it can be expressed as a matrix product state of that bond dimension.

\item The phase average is diagonal in the same basis, $\bar\rho_B(f)=\sum_jb(j;B,f)\ket{D^B_j}\bra{D^B_j}$.\newline \emph{Proof.} Expanding the product gives $\ket{v_{f,\phi}}^{\otimes B}=\sum_je^{ij\phi}\sqrt{b(j;B,f)}\,\ket{D^B_j}$, and the average over $\phi$ removes every term with $j\ne j'$.

\item At $f=n/L$, $d_{\rm tr}\big(\rho_B,\bar\rho_B(f)\big)\le B(B-1)/(2L)$.\newline \emph{Proof.} Write $b_j=b(j;B,f)=\binom{B}{j} f^j(1-f)^{B-j}$. The bound compares the hypergeometric weights $w_j$ with the binomial weights $b_j$, i.e.~sampling without and with replacement~\cite{Diaconis1980finite}. The two states are diagonal in the same orthonormal basis, so their trace distance is
\begin{equation}\label{eqapp:dtr_maxform}
d_{\rm tr}\big(\rho_B,\bar\rho_B(f)\big)=\frac{1}{2}\sum_{j=0}^B|w_j-b_j|=\sum_{j=0}^Bb_j\max\Big(1-\frac{w_j}{b_j},0\Big),
\end{equation}
where the second form uses that both sets of probabilities sum to one. It therefore suffices to show that $w_j$ is not much smaller than $b_j$. In terms of falling factorials $(x)_k=x(x-1)\cdots(x-k+1)$, the probabilities are $w_j=\binom{B}{j}(n)_j(L-n)_{B-j}/(L)_B$ and $b_j=\binom{B}{j}n^j(L-n)^{B-j}/L^B$, so that
\begin{equation}
\frac{w_j}{b_j}=\frac{\prod_{i=0}^{j-1}\big(1-\frac{i}{n}\big)\prod_{i=0}^{B-j-1}\big(1-\frac{i}{L-n}\big)}{\prod_{i=0}^{B-1}\big(1-\frac{i}{L}\big)}\ \ge\ 1-\binom{j}{2}\frac{1}{n}-\binom{B-j}{2}\frac{1}{L-n}.
\end{equation}
The inequality arises from the fact that the denominator is at most one, and $\prod_i(1-x_i)\ge1-\sum_ix_i$ (Weierstrass product inequality). Inserting this bound into Eq.~\eqref{eqapp:dtr_maxform} gives
\begin{equation}
d_{\rm tr}\big(\rho_B,\bar\rho_B(f)\big)\le\sum_{j=0}^Bb_j\Big[\frac{1}{n}\binom{j}{2}+\frac{1}{L-n}\binom{B-j}{2}\Big].
\end{equation}
Here $b_j$ is the probability of $j$ excitations when each of the $B$ sites is excited independently with probability $f$, and $\binom{j}{2}$ and $\binom{B-j}{2}$ count the pairs of sites that are both excited and both empty. Each of the $\binom{B}{2}$ pairs is doubly excited with probability $f^2$ and doubly empty with probability $(1-f)^2$, so the two averages are $\binom{B}{2}f^2$ and $\binom{B}{2}(1-f)^2$, leading to
\begin{equation}
d_{\rm tr}\big(\rho_B,\bar\rho_B(f)\big)\le\binom{B}{2}\Big[\frac{f^2}{n}+\frac{(1-f)^2}{L-n}\Big]=\frac{B(B-1)}{2L}.
\end{equation}
\end{enumerate}

\subsubsection{Typical states}
We also consider the marginal of a typical state at fixed filling $f$. Let $\psi$ be a Haar-random state at filling $f$, and let $\sigma_B$ be its marginal on any $B$ qubits. By the unitary invariance of the Haar measure, $|\psi\rangle\langle\psi|$ averages to the uniform mixture of all bit strings with $n$ excitations, and taking the partial trace over the other $L-B$ qubits gives
\begin{equation}\label{eq:typ-avg}
\bar\sigma_B=\sum_{j=0}^Bw_j\,\frac{\Pi_j}{\binom{B}{j}},
\end{equation}
where $w_j$ is defined in Lemma D1, and $\Pi_j$ is the projector onto the bit strings of $B$ qubits with $j$ excitations. Compared with the marginal of a Dicke state in Eq.~\eqref{eq:prot-marg}, each $\ket{D^B_j}$ is replaced by the uniform mixture of the bit strings with $j$ excitations, so that $\bar\sigma_B$ is a mixture of computational basis states.

To quantify the fluctuations of a single random realization about the average $\bar{\sigma}_B$, we use the second moment of the Haar measure, $\mathbb{E}\big[(\ket{\psi}\bra{\psi})^{\otimes2}\big]=(P_n\otimes P_n)(1+F)/\big(D(D+1)\big)$, where $P_n$ is the projector onto the span of the $D=\binom{L}{n}$ bit strings with $n$ excitations, and $F$ swaps the two copies. 
Writing the purity of the block as $\mathrm{Tr}\,\sigma_B^2=\mathrm{Tr}\big[(\ket{\psi}\bra{\psi})^{\otimes2}F_B\big]$, with $F_B$ the swap of the $B$ qubits only, and abbreviating $\beta_j=\binom{B}{j}$ and $a_j=\binom{L-B}{n-j}$ (so that $w_j=\beta_ja_j/D$), we find 
$\mathbb{E}\big[\mathrm{Tr}\,\sigma_B^2\big]=\sum_ja_j\beta_j(a_j+\beta_j)/\big(D(D+1)\big)$. The two terms come from $F_B$ and $FF_B$, which count the pairs of bit strings with $n$ excitations that agree on the block ($\sum_j\beta_ja_j^2$) and on its complement ($\sum_j\beta_j^2a_j$), respectively. 
This leads to
\begin{equation}\label{eq:typ-var}
\mathbb{E}\big[\|\sigma_B-\bar\sigma_B\|_2^2\big]=\mathbb{E}\big[\,\mathrm{Tr}\,\sigma_B^2\big]-\mathrm{Tr}\,\bar\sigma_B^2
=\sum_{j=0}^B\frac{a_j\beta_j(D\beta_j-a_j)}{D^2(D+1)}\ \le\ \frac{1}{D+1}\binom{B}{\lfloor B/2\rfloor},
\end{equation}
where the bound uses $\sum_jw_j=1$ and $\beta_j\le\binom{B}{\lfloor B/2\rfloor}$. Since $\sigma_B-\bar\sigma_B$ acts on a space of dimension $2^B$, its trace norm obeys $\|\sigma_B-\bar\sigma_B\|_1^2\le2^B\|\sigma_B-\bar\sigma_B\|_2^2$, and Markov's inequality gives
\begin{equation}\label{eq:typ-conc}
\Pr\big[\|\sigma_B-\bar\sigma_B\|_1\ge\varepsilon\big]\ \le\ \frac{4^B}{(D+1)\,\varepsilon^2}.
\end{equation}
At fixed filling, $D\ge2^{Lh_2(f)}/(L+1)$, so that with e.g.~$\varepsilon=2^{-Lh_2(f)/3}$ each block is exponentially close to a mixture of stabilizer states, except with a probability that is also exponentially small in $L$.

If the marginal were exactly $\bar\sigma_B$, the block would be a random bit string, i.e., a random stabilizer state. Free operations map stabilizer states to each other, so no run could output an exact CCZ state; indeed every output would have fidelity at most $9/16$ with $\ket{\CCZ}$~\cite{Bravyi2019simulation}. The actual marginal differs from $\bar\sigma_B$ by at most $\varepsilon$ in trace norm. This does not strictly constrain the output of a given run, since a rare run can in principle produce anything. However,  it constrains the output averaged over runs. Let run $r$ occur with probability $p_r$ and output a state of fidelity $F_r$ with $\ket{\CCZ}$, both evaluated on the actual marginal $\sigma_B$. For a fixed set $\mathcal R$ of runs, the sum $\sum_{r\in\mathcal R}p_r\big(F_r-\frac{9}{16}\big)$ depends linearly on the marginal, since it equals $\mathrm{Tr}(X_{\mathcal R}\sigma_B)$ with
\begin{equation}
X_{\mathcal R}=\sum_{r\in\mathcal R}\bra{0\cdots0}K_r^\dagger\Big(\ket{\CCZ}\bra{\CCZ}\otimes I-\tfrac{9}{16}\Big)K_r\ket{0\cdots0},
\end{equation}
in terms of the Kraus operators $K_r$ of App.~\ref{subsecapp:prot-proc}. Since the probabilities of all runs add up to one, $X_{\mathcal R}$ has operator norm at most one. Moreover, $\mathrm{Tr}(X_{\mathcal R}\bar\sigma_B)\le0$, because every run has fidelity at most $9/16$ on a mixture of stabilizer states. Therefore, $\mathrm{Tr}(X_{\mathcal R}\sigma_B)\le\mathrm{Tr}(X_{\mathcal R}\bar\sigma_B)+\|\sigma_B-\bar\sigma_B\|_1\le\varepsilon$. Taking for $\mathcal R$ the runs whose outputs have fidelity above $9/16+\delta$ (including all runs that output an exact CCZ state, i.e.~$\delta=7/16$) gives $\delta\sum_{r\in\mathcal R}p_r\le\varepsilon$, so these runs have a total probability of at most $\varepsilon/\delta$. Summing over the blocks, with $B$ fixed as $L\to\infty$, procedures acting on the blocks separately therefore extract on average only an exponentially small number of CCZ states.

\subsection{Block procedures}\label{subsecapp:prot-proc}

This subsection discusses how the procedures $\Lambda_6$ and $\Lambda_8$ of Ref.~\cite{Rizzo2026universal} can convert blocks of a Dicke state into CCZ states. We describe $\Lambda_6$ in full, as its three steps illustrate how such a procedure works, and then $\Lambda_8$, which has a greater yield per qubit and is the procedure used in Results 2 and 3.

A run is defined as one execution of a procedure, including its random choices and measurement outcomes. On a fixed run, we set aside the qubits that the procedure discards instead of tracing them out immediately. This means that the run applies a product of Clifford unitaries, stabilizer ancilla preparations and Pauli projections $(I\pm P)/2$, followed by a partial trace over all the qubits set aside. We consider a heralded procedure that acts on one block of $B$ qubits, together with stabilizer ancillas. Each run $r$ thus applies a single linear map $K_r$ (its Kraus operator) to the block and the ancillas, and a run is denoted as accepting if the procedure announces success. Following Ref.~\cite{Rizzo2026universal} (where the definition is stated for product inputs $\ket{\psi}^{\otimes B}$, which span $\mathrm{Sym}^B$), we call the procedure \emph{universal} if every accepting run outputs an exact CCZ state on every input supported on the symmetric subspace, i.e.,
\begin{equation}\label{eq:prot-exact}
K_r\big(\ket{v}\otimes\ket{0\cdots0}\big)=\ket{\CCZ}\otimes\ket{\zeta_r(v)}\qquad\text{for every}\ \ket{v}\in\mathrm{Sym}^B,
\end{equation}
with the CCZ state on the output qubits of the run and a vector $\ket{\zeta_r(v)}$ of the remaining qubits, which depends linearly on $\ket{v}$ and may vanish.

Since $\ket{\zeta_r(v)}$ is linear in $\ket{v}$, Eq.~\eqref{eq:prot-exact} also applies when the block is entangled with other qubits, as long as the block marginal $\rho_B$ is supported on $\mathrm{Sym}^B$. Indeed, the global state is then a sum $\sum_k\ket{v_k}\otimes\ket{\chi_k}$ with $\ket{v_k}\in\mathrm{Sym}^B$, which $K_r$ maps to $\ket{\CCZ}\otimes\sum_k\ket{\zeta_r(v_k)}\otimes\ket{\chi_k}$. The output qubits are therefore in $\ket{\CCZ}$, in tensor product with all other qubits. By Lemma D1(i), this is the case for every block of a Dicke state, and it remains so even after procedures have run on other blocks; operations on other qubits, and the conditioning on their outcomes, never take the state of a block out of the support of its marginal, since they act only on the other qubits. (The successes of different blocks are, however, correlated; see App.~\ref{subsecapp:prot-stats}.) The procedures $\Lambda_6$ and $\Lambda_8$ described below are universal \cite{Rizzo2026universal}.

The success probability of the procedure on a state $\rho$ of a block is
\begin{equation}\label{eq:prot-E}
\Pr[\text{success}]=\mathrm{Tr}\big(E\rho\big),\qquad E=\sum_{r\ \text{accepting}}\bra{0\cdots0}K_r^\dagger
K_r\ket{0\cdots0},
\end{equation}
where the success operator $E$ acts on the block alone. For a block of a Dicke state, the probability is $\sum_jw_j\bra{D^B_j}E\ket{D^B_j}$. Up to the trace distance of Lemma D1(iii), this is equal to the average over $\phi$ of the success probability on the products $\ket{v_{f,\phi}}^{\otimes B}$.

Consider the subspace of symmetric states orthogonal to the six tensor products $\ket{s}^{\otimes B}$ of a single-qubit stabilizer state,
\begin{equation}\label{eq:prot-MB}
\mathcal M_B=\big\{\ket{\Phi}\in\mathrm{Sym}^B:\ \braket{s^{\otimes B}|\Phi}=0\ \ \text{for}\,\, \ket{s} \in\{\ket{0},\ket{1},\ket{\pm},\ket{\pm i}\}\big\}.
\end{equation}
Only such components can lead to a successful run~\cite{Rizzo2026universal}. Indeed, a universal procedure must annihilate the six stabilizer tensor products (a stabilizer operation cannot turn a stabilizer state into a CCZ state), so the success operator restricted to $\mathrm{Sym}^B$ acts only on $\mathcal M_B$. Ref.~\cite{Rizzo2026universal} shows that $\mathcal M_B=\{0\}$ for $B\le5$, while $\mathcal M_B$ has dimension $B-5$ for $B\ge6$.

\subsubsection{The six-qubit procedure $\Lambda_6$}\label{subsubsecapp:six_qubit}

For six qubits, $\mathcal M_6$ is spanned by the single state $\ket{A_6}=(\ket{D^6_1}-\ket{D^6_5})/\sqrt2$~\cite{Rizzo2026universal}. The procedure $\Lambda_6$ of Ref.~\cite{Rizzo2026universal} consists of the following three steps:
\begin{enumerate}
\item Measure the parities $Z^{\otimes6}$ and $X^{\otimes6}$, and accept only the outcome $(-1,-1)$. Note that on the Dicke basis, we have
\begin{equation}\label{eq:prot-par6}
Z^{\otimes6}\ket{D^6_j}=(-1)^j\ket{D^6_j},\qquad X^{\otimes6}\ket{D^6_j}=\ket{D^6_{6-j}}.
\end{equation}
The outcome $-1$ of $Z^{\otimes6}$ keeps the span of $\ket{D^6_1}$, $\ket{D^6_3}$ and $\ket{D^6_5}$, and the outcome $-1$ of $X^{\otimes6}$ then retains only the combination that changes sign when $\ket{D^6_1}$ and $\ket{D^6_5}$ are exchanged. This step therefore projects onto $\ket{A_6}$ with probability $\bra{A_6}\rho\ket{A_6}$.
\item Measure $Z$ on qubits 5 and 6, and accept the outcomes $00$ and $11$, which occur with total probability $\frac{2}{3}$ on $\ket{A_6}$. This leaves qubits $1,\dots,4$ in the $W$ state $\ket{D^4_1}$, or in $\ket{D^4_3}=X^{\otimes4}\ket{D^4_1}$ (which can be mapped to $\ket{D^4_1}$ by applying $X^{\otimes4}$).
\item Apply a Clifford unitary that converts $\ket{D^4_1}$ to $\ket{\CCZ}\otimes\ket{0}$. This can be achieved by applying the gates $\mathrm{CNOT}_{1\to4}$, $\mathrm{CNOT}_{2\to4}$, $\mathrm{CNOT}_{3\to4}$, $X_4$, $\mathrm{CNOT}_{3\to1}$, $\mathrm{CNOT}_{3\to2}$ and $H_3$ sequentially.  Note that Ref.~\cite{Rizzo2026universal} uses an equivalent decoding with one more Pauli measurement.
\end{enumerate}
The success operator of $\Lambda_6$ on $\mathrm{Sym}^6$ is therefore $E_{\Lambda_6}=\frac{2}{3}\ket{A_6}\bra{A_6}=\frac{2}{3}\Pi_{\mathcal M_6}$, and each accepting run maps the whole symmetric subspace onto one output state, so that $\Lambda_6$ satisfies Eq.~\eqref{eq:prot-exact}.

For six copies of a single-qubit state $\ket{\psi}=c_0\ket{0}+c_1\ket{1}$, the expansion $\ket{\psi}^{\otimes6}=\sum_jc_0^{6-j}c_1^j\binom{6}{j}^{1/2}\ket{D^6_j}$ gives the overlap $\braket{A_6|\psi^{\otimes6}}=\sqrt3\,c_0c_1(c_0^4-c_1^4)$, so that $\Lambda_6$ succeeds with probability $2|c_0|^2|c_1|^2|c_0^4-c_1^4|^2$. This is a third of the linear stabilizer entropy of $\ket{\psi}$, because of the identity
\begin{equation}\label{eq:prot-mlinc}
\Mlin(\psi)=\frac{1}{2}\big(1-r_x^6-r_y^6-r_z^6\big)=6|c_0|^2|c_1|^2\big|c_0^4-c_1^4\big|^2,
\end{equation}
where $\mathbf{r}$ is the Bloch vector of $\ket{\psi}$. The second equality can be verified by writing $c_0=\sqrt{1-p}$ and $c_1=\sqrt pe^{i\phi}$, so that $r_z=1-2p$ and $r_x+ir_y=2\sqrt{p(1-p)}\,e^{i\phi}$. Eq.~\eqref{eq:prot-mlinc} can then be written as
\begin{equation}\label{eq:prot-mlinp}
6p(1-p)\big|(1-p)^2-p^2e^{4i\phi}\big|^2=6p(1-p)\big[(1-p)^4+p^4-2p^2(1-p)^2\cos4\phi\big].
\end{equation}
At latitude $p=f$, the azimuthal average of $\Mlin$ is $g(f)=6f(1-f)[(1-f)^4+f^4]$.

For a six-qubit block of $\ket{D^L_n}$ with $L\ge6$, Lemma D1(i) gives $\bra{A_6}\rho_6\ket{A_6}=\frac{1}{2}(w_1+w_5)$, with $w_j=\binom{6}{j}\binom{L-6}{n-j}/\binom{L}{n}$, so that $\Lambda_6$ succeeds with probability
\begin{equation}\label{eq:prot-p6}
p_6(L,n)=\frac{w_1+w_5}{3}=\frac{2\big[\binom{L-6}{n-1}+\binom{L-6}{n-5}\big]}{\binom{L}{n}}.
\end{equation}

\subsubsection{The eight-qubit procedure $\Lambda_8$}\label{subsubsecapp:eight_qubit}

For eight qubits, $\mathcal M_8$ is three-dimensional, with the orthonormal basis~\cite{Rizzo2026universal}
\begin{equation}\label{eq:prot-M8}
\ket{R_0}=\sqrt{\tfrac{7}{8}}\ket{D^8_1}-\sqrt{\tfrac{1}{8}}\ket{D^8_5},\qquad
\ket{R_1}=\tfrac{1}{\sqrt2}\big(\ket{D^8_2}-\ket{D^8_6}\big),\qquad
\ket{R_2}=\sqrt{\tfrac{1}{8}}\ket{D^8_3}-\sqrt{\tfrac{7}{8}}\ket{D^8_7}.
\end{equation}
The procedure $\Lambda_8$ of Ref.~\cite{Rizzo2026universal} has four steps:
\begin{enumerate}
\item Measure the parities $X^{\otimes8}$ and $Z^{\otimes8}$, which act on the Dicke basis analogously to Eq.~\eqref{eq:prot-par6}, and reject the outcome $(X^{\otimes8},Z^{\otimes8})=(+1,+1)$.
\item Map the three accepted outcomes to $(+1,-1)$ by a Clifford unitary. This can be achieved by applying $(S^\dagger)^{\otimes8}$ to $(-1,-1)$, or $H^{\otimes8}$ to $(-1,+1)$.
\item Measure $Z$ on qubit 5 and accept either outcome. Measure $X$ on each of qubits 1 to 4, and accept when the product of the four outcomes is $-1$.
\item Apply a Pauli correction and a Clifford unitary, which output $\ket{\CCZ}$.
\end{enumerate}
The success operator of $\Lambda_8$ on the symmetric subspace is~\cite{Rizzo2026universal}
\begin{equation}\label{eq:prot-succ8}
E_{\Lambda_8}=\tfrac{4}{7}\,\Pi_{\mathcal M_8}.
\end{equation}
Moreover, every accepting run outputs an exact $\ket{\CCZ}$ on every input supported on $\mathrm{Sym}^8$, pure or mixed \cite{Rizzo2026universal}.

\medskip\noindent\emph{Lemma D2 (success probabilities of $\Lambda_8$).}
\begin{enumerate}\renewcommand{\labelenumi}{(\roman{enumi})}
\item For eight copies of a single-qubit state $\ket{\psi}=c_0\ket{0}+c_1\ket{1}$, the procedure $\Lambda_8$ succeeds with probability
\begin{equation}\label{eq:prot-p8prod}
4|c_0|^2|c_1|^2\big|c_0^4-c_1^4\big|^2=\tfrac{2}{3}\Mlin(\psi).
\end{equation}
For $\ket{v_{f,\phi}}^{\otimes8}$, this equals
\begin{equation}\label{appeq:mathcalP_f_phi}
\mathcal{P}_f(\phi)=4f(1-f)\,\big|(1-f)^2-f^2e^{4i\phi}\big|^2.
\end{equation}
\emph{Proof.} Expanding $\ket{\psi}^{\otimes8}=\sum_jc_0^{8-j}c_1^j\binom{8}{j}^{1/2}\ket{D^8_j}$ and taking the overlaps with Eq.~\eqref{eq:prot-M8} gives
\begin{equation}
\braket{R_0|\psi^{\otimes8}}=\sqrt7\,c_0^3c_1\,(c_0^4-c_1^4),\qquad
\braket{R_1|\psi^{\otimes8}}=\sqrt{14}\,c_0^2c_1^2(c_0^4-c_1^4),\qquad
\braket{R_2|\psi^{\otimes8}}=\sqrt7\,c_0c_1^3(c_0^4-c_1^4).
\end{equation}
The squared norm of the projection onto $\mathcal M_8$ is therefore $7|c_0|^2|c_1|^2|c_0^4-c_1^4|^2$, and Eq.~\eqref{eq:prot-succ8} gives the LHS of Eq.~\eqref{eq:prot-p8prod}, which is $\frac{2}{3}\Mlin(\psi)$ by Eq.~\eqref{eq:prot-mlinc}. For $c_0=\sqrt{1-f}$ and $c_1=\sqrt fe^{i\phi}$, it is $4f(1-f)|(1-f)^2-f^2e^{4i\phi}|^2=\mathcal{P}_f(\phi)$.
\item For an eight-qubit block of $\ket{D^L_n}$, $\Lambda_8$ succeeds with probability
\begin{equation}\label{eq:prot-p8dicke}
p_8(L,n)=\frac{4\big[\binom{L-6}{n-1}+\binom{L-6}{n-5}\big]}{\binom{L}{n}}=2p_6(L,n).
\end{equation}
\emph{Proof.} By Lemma D1(i) and Eq.~\eqref{eq:prot-succ8}, the success probability is $\frac{4}{7}\sum_jw_j\bra{D^8_j}\Pi_{\mathcal M_8}\ket{D^8_j}$. Eq.~\eqref{eq:prot-M8} gives the diagonal elements $\frac{7}{8}$ for $j=1,7$, $\frac{1}{2}$ for $j=2,6$, $\frac{1}{8}$ for $j=3,5$ and $0$ otherwise. With $w_j=\binom{8}{j}\binom{L-8}{n-j}/\binom{L}{n}$,
\begin{equation}
\binom{L}{n}\,p_8(L,n)=4\Big[\tbinom{L-8}{n-1}+2\tbinom{L-8}{n-2}+\tbinom{L-8}{n-3}\Big]
+4\Big[\tbinom{L-8}{n-5}+2\tbinom{L-8}{n-6}+\tbinom{L-8}{n-7}\Big],
\end{equation}
and the identity $\binom{L-6}{a}=\binom{L-8}{a}+2\binom{L-8}{a-1}+\binom{L-8}{a-2}$ gives Eq.~\eqref{eq:prot-p8dicke}.
\end{enumerate}

\medskip\noindent\emph{The four-qubit $W$ state.} For $L=4$ and $n=1$, no procedure is needed, since $\ket{D^4_1}$ is Clifford-equivalent to $\ket{\CCZ}\otimes\ket{0}$. With the qubits numbered $1,\dots,4$, applying $H_3$, $\mathrm{CNOT}_{3\to2}$, $\mathrm{CNOT}_{3\to1}$, $X_4$, $\mathrm{CNOT}_{3\to4}$, $\mathrm{CNOT}_{2\to4}$ and $\mathrm{CNOT}_{1\to4}$ to $\ket{\CCZ}\otimes\ket{0}$ gives $\ket{D^4_1}$. Hence $\ket{D^4_1}$ applies one Toffoli gate deterministically.

\subsection{Statistics of the blockwise yield}\label{subsecapp:prot-stats}

In this subsection, we derive the limiting distribution of the number of extracted CCZ states, which proves Results 2 and 3. The idea is that the fraction $t/N$ of successful blocks behaves like the success probability of one block with a random hidden phase. 
Each block of a Dicke state is close to the phase average of products of tilted spins (Lemma D1), and all blocks share a common hidden phase. We first derive this distribution for any fixed block procedure, and then specialize to $\Lambda_8$ (we consider $\Lambda_8$ instead of $\Lambda_6$ because it has a greater yield per qubit).

\medskip\noindent\emph{Proposition D3 (distribution of the success fraction).} Let $E$ be the success operator of some fixed procedure, i.e., the sum of $K_r^\dagger K_r$ over its accepting runs. Apply it independently to each of the $N=\lfloor L/B\rfloor$ disjoint blocks of $\ket{D^L_n}$, and let $t$ be the number of successful blocks. For $L\rightarrow\infty$, the probability distribution for the success fraction $t/N$ converges to that of
\begin{equation}\label{eq:prot-qf}
q_f(\Phi)=\bra{v_{f,\Phi}}^{\otimes B}E\ket{v_{f,\Phi}}^{\otimes B},
\end{equation}
with $\Phi$ uniform on $[0,2\pi)$.

\medskip\noindent\emph{Proof.} We first show that the moments of $t/N$ converge to those of $q_f(\Phi)$, and then pass from the moments to the distribution. The procedures act on disjoint qubits, with no feed-forward between blocks, so the probability that all blocks of a set $\mathcal T$ of $u$ blocks succeed is the expectation value of $\bigotimes_{i\in\mathcal T}E_i$ in the Dicke state, with $E_i$ acting on block $i$. By permutation symmetry, the marginal of $\ket{D^L_n}$ on these $uB$ qubits is $\rho_{uB}$ for any set $\mathcal{T}$, so that
\begin{equation}\label{eq:prot-mu}
\Pr[\text{all blocks of }\mathcal T\text{ succeed}]=\mu_u:=\mathrm{Tr}\big(\rho_{uB}E^{\otimes u}\big).
\end{equation}
Let $\chi_i=1$ if the procedure succeeds on block $i$, and $\chi_i=0$ otherwise, so that $t=\sum_i\chi_i$. Picking $u$ different successful blocks in order can be done in $t(t-1)\cdots(t-u+1)$ ways, and the product $\chi_{i_1}\cdots\chi_{i_u}$ is one exactly when the blocks $i_1,\dots,i_u$ all succeed, so that
\begin{equation}
t(t-1)\cdots(t-u+1)=\sum_{i_1,\dots,i_u\ \text{distinct}}\chi_{i_1}\cdots\chi_{i_u}.
\end{equation}
By Eq.~\eqref{eq:prot-mu}, each of the $N(N-1)\cdots(N-u+1)$ terms on the RHS above has expectation $\mu_u$, so that the factorial moments of $t$ are
\begin{equation}\label{eqapp:factorial_moment_t}
\mathbb E\big[t(t-1)\cdots(t-u+1)\big]=N(N-1)\cdots(N-u+1)\,\mu_u .
\end{equation}

To determine $\mu_u$, Lemma D1(iii) applied to $uB$ qubits gives (with $0\le E^{\otimes u}\le I$)
\begin{equation}
\Big|\mu_u-\mathrm{Tr}\big(\bar\rho_{uB}(f)E^{\otimes u}\big)\Big|\le d_{\rm tr}\big(\rho_{uB},\bar\rho_{uB}(f)\big)
\le\frac{uB(uB-1)}{2L}.
\end{equation}
The bound vanishes as $L\to\infty$ for fixed $u$, so that $\mu_u$ may be evaluated in the phase-averaged state, where all $uB$ qubits are in $\ket{v_{f,\phi}}$ with one common phase $\phi$ drawn uniformly from $[0,2\pi)$. For a fixed $\phi$, the $u$ blocks are in the product state $\big(\ket{v_{f,\phi}}^{\otimes B}\big)^{\otimes u}$, in which the expectation value of $E^{\otimes u}$ factorizes into $u$ copies of the single-block success probability $q_f(\phi)$ of Eq.~\eqref{eq:prot-qf}. Averaging over $\phi$ gives
\begin{equation}
\mathrm{Tr}\big(\bar\rho_{uB}(f)E^{\otimes u}\big)=\int_0^{2\pi}\frac{d\phi}{2\pi}\,q_f(\phi)^u=\mathbb E\big[q_f(\Phi)^u\big]
=:m_u,
\end{equation}
the $u$-th moment of $q_f(\Phi)$. Note that due to the common phase, the result is the average of $q_f^u$, rather than the $u$-th power of the average of $q_f$. Since $q_f(\phi)$ is a polynomial in $\sqrt f$, $\sqrt{1-f}$ and $e^{\pm i\phi}$, $m_u$ depends continuously on $f$, and $\mu_u\to m_u$ as $L\to\infty$.

We now convert from the factorial moments (Eq.~\eqref{eqapp:factorial_moment_t}) to the ordinary moments of $t/N$. The difference $t^u-t(t-1)\cdots(t-u+1)$ is a polynomial of degree $u-1$ in $t$, with coefficients that depend only on $u$. Since $0\le t\le N$, the difference is $O(N^{u-1})$. Dividing the factorial moments by $N^u$ therefore gives
\begin{equation}
\mathbb E\big[(t/N)^u\big]=\frac{N(N-1)\cdots(N-u+1)}{N^u}\,\mu_u+O(1/N)\ \longrightarrow\ m_u
\end{equation}
for every $u$. Since $t/N$ and $q_f(\Phi)$ both lie in a bounded range $[0,1]$, matching moments correspond to matching distributions, so that $t/N\to q_f(\Phi)$ in distribution. $\square$

We now apply $\Lambda_8$ to the $N=\lfloor L/8\rfloor$ blocks. Each successful block outputs one exact CCZ state (App.~\ref{subsecapp:prot-proc}), so the CCZ yield is the number $t$ of successful blocks.

\subsubsection{Average yield}

The mean of $t$ is $\mathbb E[t]=N\,p_8(L,n)$ by Lemma D2(ii). For fixed $j$, the ratio $\binom{L-6}{n-j}/\binom{L}{n}=(n)_j(L-n)_{6-j}/(L)_6$, with the falling factorials defined in the proof of Lemma D1(iii), tends to $f^j(1-f)^{6-j}$. Hence
\begin{equation}
p_8(L,n)\to4\big[f(1-f)^5+f^5(1-f)\big]=\tfrac{2}{3}g(f),
\end{equation}
and with $N/L\to\frac{1}{8}$ this gives the average CCZ yield of $\mathbb E[t]/L\to g(f)/12$. Equivalently, $\frac{2}{3}g(f)$ is the average of $\mathcal{P}_f(\Phi)$ over $\Phi$ (see Eq.~\eqref{appeq:mathcalP_f_phi}).

\subsubsection{Distribution of the yield}

By Lemma D2(i), Eq.~\eqref{eq:prot-qf} for $\Lambda_8$ is $q_f=\mathcal{P}_f$, so Proposition D3 gives $t/N\to\mathcal{P}_f(\Phi)$ in distribution. For $0<f<1$ the coefficient $-8f^3(1-f)^3$ of $\cos4\Phi$ in $\mathcal{P}_f(\Phi)$ does not vanish, so $t/L$ does not concentrate (i.e.~non-self-averaging).

\subsubsection{Guaranteed rate}

Recall that the guaranteed rate is the largest $r$ such that at least $rL$ CCZ states are obtained with a probability tending to one (Sec.~\ref{subsec:ccz}). The minimum of $\mathcal{P}_f(\Phi)$ over $\Phi$ lies at $\Phi\in\frac{\pi}{2}\mathbb Z$, where it equals
\begin{equation}
x_{\min}=4f(1-f)(1-2f)^2 .
\end{equation}
The guaranteed rate is set by the least favourable hidden phase. By Result 3, the fraction $t/N$ of successful blocks behaves for large $L$ like $\mathcal{P}_f(\Phi)$, which is at least $x_{\min}$ for every $\Phi$. For every $x<x_{\min}$, the probability that $t/N\ge x$ therefore tends to one. For $x>x_{\min}$ it does not, since $\Phi$ lies close to $\frac{\pi}{2}\mathbb Z$ with positive probability. With $L/N\to8$, the guaranteed rate is therefore $x_{\min}/8=\frac{1}{2}f(1-f)(1-2f)^2$, which vanishes at half-filling.

\clearpage
\newpage

\section{From CCZ to $T$ states}\label{app:T}

\subsection{No exact $T$ state without a catalyst}\label{subsecapp:T-field}

We first demonstrate that it is impossible to extract exact $T$ states without a catalyst, using the number-field constraints of Ref.~\cite{Beverland2019lower}. For a number field that contains $\QI$ (the complex numbers with rational real and imaginary parts, i.e.~Gaussian rationals) and is closed under complex conjugation, a stabilizer circuit (Clifford unitaries, ancillas in $\ket{0}$ and Pauli measurements, with or without post-selection) maps every density matrix whose entries lie in the field, to another whose entries lie in the same field. For the field  $\QI$, we apply this idea to each run of a procedure without a catalyst. Note that discarding qubits takes a partial trace, which only sums entries. Intuitively, no step leaves $\QI$, because conjugation by a Hadamard gate only introduces a factor $\frac{1}{2}$ and signs, the other Clifford gates have entries $0$, $\pm1$ and $\pm i$, and a Pauli measurement applies a projector $(I\pm P)/2$ and divides by the outcome probability, which is rational.

Dicke states have density matrices with the entries $0$ and $1/\binom{L}{n}$, so the output of any procedure has all entries in $\QI$. In contrast, the density matrix of the $T$ state $\ket{T}=(\ket{0}+e^{i\pi/4}\ket{1})/\sqrt2$ has the off-diagonal entry $e^{-i\pi/4}/2=(1-i)/(2\sqrt2)$, which is not in $\QI$. If one qubit of the output were exactly in such a state, discarding all other qubits would yield a run whose output is this state, which is impossible. Hence, a procedure without a catalyst cannot output a qubit that is exactly in a $T$ state, starting from Dicke states.

\subsection{$T$ yields with one $T$ state as a catalyst}\label{subsecapp:T-yields}

The obstruction above disappears if a $T$ state may be borrowed as a catalyst. Ref.~\cite{Gidney2018efficient} provides a circuit of Clifford gates and a single $T$ gate that converts $\ket{\CCZ}$ exactly into $\ket{T}^{\otimes3}$. The $T$ gate is applied by gate injection \cite{Gottesman1999demonstrating,Zhou2000methodology}, which consumes one $T$ state. One of the three output $T$ states then serves as the catalyst of the next conversion, so that a chain of $t$ conversions turns $t$ CCZ states and one catalyst $T$ state into $2t+1$ $T$ states.

The catalysed $T$ yield from Dicke states straightforwardly follows by converting the CCZ outputs of $\Lambda_8$. The procedure runs $\Lambda_8$ on each of the $N=\lfloor L/8\rfloor$ blocks of $\ket{D^L_n}$, as for Results 2 and 3, and then applies the catalysis chain above to the $t$ CCZ states. This outputs $2t$ $T$ states and returns the catalyst exactly on every run. The overall procedure is therefore heralded with one catalyst, and its $T$ count $t_T=2t$ is twice the CCZ count of Results 2 and 3. From App.~\ref{subsecapp:prot-stats}, we deduce that (see Eq.~\eqref{eq:ccz})
\begin{equation}\label{eq:tyield}
\lim_{L\to\infty}\frac{\mathbb E[t_T]}{L}=\frac{g(f)}{6}=f(1-f)\big[(1-f)^4+f^4\big],
\end{equation}
which is $1/32$ at half-filling. Furthermore, every $T$-yield rate below $f(1-f)(1-2f)^2$ (which vanishes at half-filling) is guaranteed.

\clearpage
\newpage
\section{Bounds on cost and yield}\label{app:upper}

In this appendix, we prove the lower bound on the cost of preparing a Dicke state, discuss a heralded preparation whose cost lies within a factor of about three of the bound, and prove the upper bounds on the $T$ and CCZ yields (for any run) and the bound on the relative entropy of magic (which constrains averages over runs).

\subsection{Pauli symmetries of a Dicke state}\label{subsecapp:up-pauli}

We first discuss the Pauli symmetries of a Dicke state, which will enter the single-run bounds in App.~\ref{subsecapp:up-cost}. For a pure state $\ket{\psi}$ of $L$ qubits, let $\mathrm{Stab}\,\psi$ be the group of signed Pauli strings $\pm P$ that leave $\ket{\psi}$ invariant. Its stabilizer nullity \cite{Beverland2019lower}
\begin{equation}\label{eq:up-nu}
\nu(\psi)=L-\log_2|\mathrm{Stab}\,\psi|
\end{equation}
is a logarithmic count of the missing Pauli symmetries. $\nu(\psi)$ is invariant under Clifford unitaries, additive under tensor products and zero exactly on stabilizer states, and it does not increase on any outcome of a Pauli measurement that occurs with non-zero probability~\cite{Beverland2019lower}. A $T$ state has $\nu=1$ and a CCZ state has $\nu=3$ (the maximum possible values for 1- and 3-qubit states respectively).

For Dicke states with $1\le n\le L-1$, it can be shown that
\begin{equation}
\mathrm{Stab}\,D^L_n=\begin{cases}\{I,(-1)^nZ^{\otimes L}\},\quad \nu(D^L_n)=L-1, & 2n\ne L,\\[2pt]
\{I,X^{\otimes L},Y^{\otimes L},(-1)^{L/2}Z^{\otimes L}\},\quad \nu(D^L_n)=L-2, & 2n=L.\end{cases}
\end{equation}
Indeed, up to a phase, every Pauli string is $X^aZ^z$ with bit strings $a$ and $z$, which maps $\ket{x}$ to $(-1)^{z\cdot x}\ket{x\oplus a}$. To leave $\ket{D^L_n}$ invariant up to a sign, it must map the strings of weight $n$ onto themselves, which forces $a=0^L$ (at half-filling, we have the additional option $a=1^L$), and it must give all of them the same sign, which forces $z=0^L$ or $z=1^L$.

\subsection{Bounds from stabilizer nullity}\label{subsecapp:up-cost}

Since free operations cannot raise the nullity, the nullity of the resource states bounds the nullity of the output of any run performed on such states. This furnishes both the cost bound, and the single-run bounds on the yields.

\medskip\noindent\emph{Proposition F1 (cost and yields on every run).} For heralded procedures, with or without a pure catalyst:
\begin{enumerate}
\item[(i)] every run that outputs $\ket{D^L_n}$ exactly consumes at least $\nu(D^L_n)$ $T$ states, or at least $\nu(D^L_n)/3$ CCZ states;
\item[(ii)] every run that extracts $t$ CCZ states exactly from $\ket{D^L_n}$ has $t\le\nu(D^L_n)/3$, and every run that extracts $t$ $T$ states exactly has $t\le\nu(D^L_n)$.
\end{enumerate}

\medskip\noindent\emph{Proof.} On a fixed run, we set aside the qubits that the procedure discards instead of tracing them out, so that the joint state of all qubits stays pure. For (i), let the run act on $t$ $T$ states and a pure catalyst $\ket{\varphi}$. Since stabilizer ancillas have zero nullity, the initial nullity is $t+\nu(\varphi)$, which is not increased by Clifford unitaries or measurement outcomes. At the end, the kept registers are exactly in $\ket{D^L_n}\otimes\ket{\varphi}$. Since a pure state with a pure marginal is a product state, the joint state is $\ket{D^L_n}\otimes\ket{\varphi}\otimes\ket{\chi}$, with $\ket{\chi}$ the state of the qubits set aside. Additivity gives
\begin{equation}
\nu(D^L_n)+\nu(\varphi)+\nu(\chi)\le t+\nu(\varphi),
\end{equation}
so that $t\ge\nu(D^L_n)$. With $t$ CCZ states, the initial nullity is $3t+\nu(\varphi)$, which gives $3t\ge\nu(D^L_n)$. For (ii), the inputs and outputs are exchanged: the initial nullity is $\nu(D^L_n)+\nu(\varphi)$, and the final state is $\ket{\CCZ}^{\otimes t}\otimes\ket{\varphi}\otimes\ket{\chi}$, so that $3t+\nu(\varphi)\le\nu(D^L_n)+\nu(\varphi)$. Similarly, for $t$ $T$ states, we have $t+\nu(\varphi)\le\nu(D^L_n)+\nu(\varphi)$. $\square$

\subsection{A heralded preparation}\label{subsecapp:up-prep}

\medskip\noindent\emph{Proposition F2 (heralded preparation).} There is a heralded procedure that consumes $L-s_2(L)$ CCZ states, or $4(L-s_2(L))$ $T$ states, where $s_2(L)$ is the number of ones in the binary expansion of $L$. Acting on $\ket{+}^{\otimes L}$ and stabilizer ancillas, it outputs $\ket{D^L_w}$ (with all ancillas returned to $\ket{0}$) with probability $\binom{L}{w}2^{-L}$.

\medskip\noindent\emph{Proof.} We follow the heralded preparation of Ref.~\cite{Yu2024efficient}, which involves measuring the number of excitations in $\ket{+}^{\otimes L}$. To do this with free operations, we compute the weight of each basis state into an ancilla register, $\ket{x}\ket{0}\mapsto\ket{x}\ket{|x|}$, which takes $L-s_2(L)$ Toffoli gates and as many clean ancillas~\cite{Campbell2021early}. 
Each Toffoli gate costs one CCZ state through gate teleportation, or four $T$ gates~\cite{Jones2012novel}. Measuring the register gives $w$ with probability $\binom{L}{w}2^{-L}$ and projects the $L$ qubits onto $\ket{D^L_w}$. 
Resetting the register and the ancillas requires only $X$ measurements and Clifford corrections~\cite{Gidney2017halving}, which leaves $\ket{D^L_w}\ket{0}$. $\square$

\subsection{Relative entropy of magic}\label{subsecapp:up-dm}

The nullity argument in App.~\ref{subsecapp:up-cost} enforces a bound on every run, but it ignores how often a successful run occurs. The relative entropy of magic~\cite{Veitch2014resource} accounts for these probabilities and bounds the \emph{average} yield. It is defined as
\begin{equation}\label{eq:up-DM}
D_M(\rho)=\min_{\sigma\in\mathrm{STAB}}D(\rho\|\sigma),
\end{equation}
with the quantum relative entropy $D(\rho\|\sigma)=\mathrm{Tr}\,\rho\,(\log_2\rho-\log_2\sigma)$, and STAB the set of mixtures of stabilizer states. We use two of its properties:
\begin{enumerate}

\item $D_M$ does not increase on average under procedures built from free operations. This includes heralded procedures: if run $r$ occurs with probability $p_r$ and leaves the normalized output $\rho_r$, then $\sum_rp_rD_M(\rho_r)\le D_M(\rho)$ \cite{Veitch2014resource,Rubboli2023additivity}.
\item $D_M(\CCZ^{\otimes t})\ge t\log_2\frac{16}{9}$~\cite{Rubboli2023additivity,Rizzo2026universal}.
\end{enumerate}
Let a heralded procedure without a catalyst extract $t_r$ CCZ states from $\ket{D^L_n}$ on run $r$, with $t_r=0$ on the failed runs, and discard all other qubits at the end of each run. The above properties give
\begin{equation}\label{eq:up-cczav}
D_M(D^L_n)\ \ge\ \sum_rp_rD_M(\CCZ^{\otimes t_r})\ \ge\ \log_2\frac{16}{9}\;\mathbb E[t].
\end{equation}
From Result 2, the heralded procedure $\Lambda_8$, applied to all $\lfloor L/8\rfloor$ blocks, extracts on average $g(f)/12$ exact CCZ states per qubit. Eq.~\eqref{eq:up-cczav} then gives, for large $L$,
\begin{equation}
D_M(D^L_n)\ \ge\ \log_2\frac{16}{9}\cdot\frac{1}{2}f(1-f)\big[(1-f)^4+f^4\big]\,L\,\big(1+o(1)\big),
\end{equation}
so that $D_M$ of a Dicke state is extensive at $0<f<1$, including at half-filling.

\clearpage
\newpage
\section{Dicke states in model Hamiltonians}\label{app:settings}

In this appendix, we show that each of the three families of towers of Sec.~\ref{sec:outlook} is the image of the Dicke states under a Clifford unitary that does not depend on the position $n$ in the tower, possibly with idle qubits in $\ket{0}$. The states of these towers therefore have the same magic, cost and yields as the Dicke states.

\subsection{Toy model of Ref.~\cite{Choi2018emergent}}\label{subsecapp:set-toy}

The toy model of Ref.~\cite{Choi2018emergent} has $L\ge4$ spin-1/2 degrees of freedom on a ring, whose Hamiltonian after a global Hadamard rotation can be expressed as
\begin{equation}\label{eq:set-toy}
H_{\rm toy}=\frac{\Omega}{2}\sum_iZ_i+\sum_iV_{i-1,i+2}\,\Pi_{i,i+1},\qquad
\Pi_{i,i+1}=\tfrac{1}{4}\big(1-\vec\sigma_i\cdot\vec\sigma_{i+1}\big),
\end{equation}
with $V_{i-1,i+2}$ an arbitrary Hermitian operator on the spins $i-1$ and $i+2$. The tower of Ref.~\cite{Choi2018emergent} consists of the states of maximal total spin $L/2$, which correspond to Dicke states. Indeed, $\Pi_{i,i+1}$ annihilates every state that is symmetric under the exchange of the spins $i$ and $i+1$, so every interaction term annihilates the Dicke states. The Zeeman term then gives
\begin{equation}\label{eq:set-scar}
H_{\rm toy}\ket{D^L_n}=\Omega\big(\tfrac{L}{2}-n\big)\ket{D^L_n},\qquad 0\le n\le L.
\end{equation}

\subsection{$\eta$-pairing}\label{subsecapp:set-eta}

We consider spin-$\frac{1}{2}$ fermions on $L$ sites $r=1,\dots,L$ of any lattice, so that the Fock space consists of the $2L$ fermionic modes $(r,\uparrow)$ and $(r,\downarrow)$. In terms of the pair operators $b^\dagger_r=c^\dagger_{r\uparrow}c^\dagger_{r\downarrow}$, the $\eta$-pairing states are
\begin{equation}
\ket{\eta_n}=\frac{(\eta^\dagger)^n\ket{\rm vac}}{\|(\eta^\dagger)^n\ket{\rm vac}\|},\qquad
\eta^\dagger=\sum_rq_r\,b^\dagger_r,\qquad q_r\in\{\pm1,\pm i\},\qquad 0\le n\le L.
\end{equation}
The $\eta$-pairing tower of the Hubbard model on a bipartite lattice has $q_r=\pm1$, with the sign set by the sublattice \cite{Yang1989eta,Moudgalya2020etapairing}. We encode the $2L$ modes into $2L$ qubits using the Jordan-Wigner encoding~\cite{Jordan1928uber} with the mode order $(1{\uparrow},1{\downarrow},2{\uparrow},\dots,L{\downarrow})$. 
This maps the Fock state $c^\dagger_{\mu_1}\cdots c^\dagger_{\mu_j}\ket{\rm vac}$, with $\mu_1<\dots<\mu_j$, to the basis state with $\ket{1}$ on the qubits of the occupied modes $\mu_1,\dots,\mu_j$ and $\ket{0}$ on all others.

The pair operators commute with each other and square to zero. Hence,
\begin{equation}
(\eta^\dagger)^n\ket{\rm vac}=n!\sum_{|A|=n}\prod_{r\in A}q_rb^\dagger_r\ket{\rm vac},
\end{equation}
where $A$ runs over the sets of $n$ sites, and the norm is $n!\binom{L}{n}^{1/2}$. 
Each term is encoded as the basis state $\ket{x(A)}$ with ones on both qubits of every site in $A$. To relate this to Dicke states, we place the Dicke qubits on the qubits $(r,\uparrow)$ and the idle qubits on $(r,\downarrow)$. A CNOT gate from $(r,\uparrow)$ to $(r,\downarrow)$ on every site copies the occupations, and the gate $\mathrm{diag}(1,q_r)$ on $(r,\uparrow)$ (supplied by $I$, $Z$, $S$ or $S^\dagger$) generates the phases. The encoded tower state is therefore
\begin{equation}
\ket{\eta_n}=\binom{L}{n}^{-1/2}\sum_{|A|=n}\Big(\prod_{r\in A}q_r\Big)\ket{x(A)}
=\bigotimes_r\mathrm{diag}(1,q_r)_{r\uparrow}\ \prod_r\mathrm{CNOT}_{r\uparrow\to r\downarrow}\,
\big(\ket{D^L_n}\otimes\ket{0}^{\otimes L}\big),
\end{equation}
where the Clifford unitary does not depend on $n$.

\subsection{Spin-1 bimagnons}\label{subsecapp:set-bimagnon}

The spin-1 XY magnets of Ref.~\cite{Schecter2019weak}, on $V$ sites $j$ of a hypercubic lattice at the positions $\mathbf r_j$, have the bimagnon tower
\begin{equation}
\ket{\mathcal S_n}\propto(J^+)^n\ket{\Omega},\qquad\ket{\Omega}=\bigotimes_j\ket{m_j=-1},\qquad
J^+=\frac{1}{2}\sum_je^{i\mathbf r_j\cdot\boldsymbol\pi}(S^+_j)^2,\qquad 0\le n\le V,
\end{equation}
with $\boldsymbol\pi=(\pi,\dots,\pi)$. Since $(S^+)^2$ maps $\ket{-1}$ to $2\ket{+1}$ and annihilates $\ket{0}$ and $\ket{+1}$, the tower only has support on the two levels $\ket{-1}$ and $\ket{+1}$ of each site, and
\begin{equation}
\ket{\mathcal S_n}=\binom{V}{n}^{-1/2}\sum_{|A|=n}\Big(\prod_{j\in A}s_j\Big)\ket{+1\ \text{on}\ A,\ -1\ \text{elsewhere}},    \qquad s_j=e^{i\mathbf r_j\cdot\boldsymbol\pi}=\pm1.
\end{equation}
Each site therefore acts as a qubit, with $\ket{-1}\rightarrow\ket{0}$ and $\ket{+1}\rightarrow\ket{1}$, and the tower state is $\ket{D^V_n}$ with a $Z$ gate applied on every site
with $s_j=-1$.

\end{document}